\documentclass[[twocolumn,traditabstract,a4]{aa}
\usepackage{times}
\usepackage{graphicx}
\usepackage{subfigure}
\usepackage{amsmath}
\usepackage{color}
\usepackage{multirow}
\usepackage{multicol}
\usepackage{float}
\usepackage{tabularx}
\usepackage{chngcntr}
\usepackage[toc,page]{appendix}
\bibpunct{(}{)}{;}{a}{}{,} 
\usepackage[colorlinks=true,linkcolor=red,citecolor=blue, breaklinks=true]{hyperref}
\usepackage{soul}
\usepackage{natbib,twoopt}
\usepackage[normalem]{ulem}
\usepackage[toc,page]{appendix}

\makeatletter
\newcommandtwoopt{\citeads}[3][][]{\href{http://adsabs.harvard.edu/abs/#3}%
{\def\hyper@linkstart##1##2{}%
\let\hyper@linkend\@empty\citealp[#1][#2]{#3}}}
\newcommandtwoopt{\citepads}[3][][]{\href{http://adsabs.harvard.edu/abs/#3}%
{\def\hyper@linkstart##1##2{}%
\let\hyper@linkend\@empty\citep[#1][#2]{#3}}}
\newcommandtwoopt{\citetads}[3][][]{\href{http://adsabs.harvard.edu/abs/#3}%
{\def\hyper@linkstart##1##2{}%
\let\hyper@linkend\@empty\citet[#1][#2]{#3}}}
\newcommandtwoopt{\citeyearads}[3][][]%
{\href{http://adsabs.harvard.edu/abs/#3}
{\def\hyper@linkstart##1##2{}%
\let\hyper@linkend\@empty\citeyear[#1][#2]{#3}}}
\makeatother

\providecommand{\sorthelp}[1]{}
\newbox\tablebox    \newdimen\tablewidth
\def\leaderfil{\leaders\hbox to 5pt{\hss.\hss}\hfil}

\def\endPlancktablewide{\tablewidth=\textwidth 
    $$\hss\copy\tablebox\hss$$
    \vskip-\lastskip\vskip -2pt}
\def\tablenote#1 #2\par{\begingroup \parindent=0.8em
    \abovedisplayshortskip=0pt\belowdisplayshortskip=0pt
    \noindent
    $$\hss\vbox{\hsize\tablewidth \hangindent=\parindent \hangafter=1 \noindent
    \hbox to \parindent{$^#1$\hss}\strut#2\strut\par}\hss$$
    \endgroup}
\def\doubleline{\vskip 3pt\hrule \vskip 1.5pt \hrule \vskip 5pt}
\def\deg{\ifmmode^\circ\else$^\circ$\fi}
\def\pdeg{\ifmmode $\setbox0=\hbox{$^{\circ}$}\rlap{\hskip.11\wd0 .}$^{\circ}
          \else \setbox0=\hbox{$^{\circ}$}\rlap{\hskip.11\wd0 .}$^{\circ}$\fi}
\def\arcm{\ifmmode {^{\scriptstyle\prime}}
          \else $^{\scriptstyle\prime}$\fi}

\def\GHz{\ifmmode $\,GHz$\else \,GHz\fi}
\def\muKs{\ifmmode \,\mu$K\,s$^{1/2}\else \,$\mu$K\,s$^{1/2}$\fi}
\def\MJysr{\ifmmode \,$MJy\,sr\mo$\else \,MJy\,sr\mo\fi}
\def\KJysr{\ifmmode \,$kJy\,sr\mo$\else \,kJy\,sr\mo\fi}
\def\MJysrmK{\ifmmode \,$MJy\,sr\mo$\,mK$_{\rm CMB}\mo\else \,MJy\,sr\mo\,mK$_{\rm CMB}\mo\fi}
\def\muK{\ifmmode \,\mu$K$\else \,$\mu$\hbox{K}\fi}
\def\microK{\ifmmode \,\mu$K$\else \,$\mu$\hbox{K}\fi}
\def\kms{\ifmmode $\,km\,s$^{-1}\else \,km\,s$^{-1}$\fi}
\def\cm2{\ifmmode $\,cm$^{-2}\else \,cm$^{-2}$\fi}
\def\kmsMpc{\ifmmode $\,\kms\,Mpc\mo$\else \,\kms\,Mpc\mo\fi}
\def\Mpc{\ifmmode $\,Mpc\mo$\else\,Mpc\mo\fi}

\DeclareMathAlphabet{\mathsc}{OT1}{cmr}{m}{sc}
\def\testbx{bx}%
\DeclareRobustCommand{\ion}[2]{%
\relax\ifmmode
\ifx\testbx\f@series
{\mathbf{#1\,\mathsc{#2}}}\else
{\mathrm{#1\,\mathsc{#2}}}\fi
\else\textup{#1\,{\mdseries\textsc{#2}}}%
\fi}

\newcommand{\planck}{{\it Planck\/}}

\newcommand{\healpix}{\ensuremath{\tt HEALPix}}
\newcommand{\cosmo}{\ensuremath{\tt COSMO}}
\newcommand{\processmask}{\ensuremath{\tt Process\char`_mask}}
\newcommand{\smica}{\ensuremath{\tt SMICA}}
\newcommand{\sevem}{\ensuremath{\tt SEVEM}}
\newcommand{\GNILC}{\ensuremath{\tt GNILC}}
\newcommand{\xpol}{\ensuremath{\tt Xpol}}

\newcommand{\kcmb}{\ensuremath{{\rm K}_{\rm CMB}}}
\newcommand{\mukcmb}{\ensuremath{{\mu \rm K}_{\rm CMB}}}
\newcommand{\mukcmbsq}{\ensuremath{{\mu \rm K}_{\rm CMB}}^2}

\newcommand{\NH}{\ensuremath{N_\mathsc {H}}}
\newcommand{\NHI}{\ensuremath{N_\mathsc {Hi}}}
\newcommand{\NHIi}{\ensuremath{N_\mathsc {Hi}^i}}
\newcommand{\hi}{\ensuremath{\mathsc {Hi}}}
\newcommand{\hii}{\ensuremath{\mathsc {Hii}}}

\newcommand{\hmol}{H$_2$}

\newcommand{\rgn}{mask65}
\newcommand{\ignilc}{\ensuremath{I_{G353}}}
\newcommand{\tgmodel}{TG17}
\def\Planck{\textit{Planck}}
\def\mo{\ifmmode^{-1}\else$^{-1}$\fi}
\def\,{\thinspace}

\newcommand{\Nside}{\ensuremath{N_{\rm side}}}

\newcommand{\Bt}{\ensuremath{\vec{B}_{\rm turb}}}
\newcommand{\Bo}{\ensuremath{\vec{B}_{\rm ord}}}
\newcommand{\BT}{\ensuremath{\vec{B}}}
\newcommand{\Bpos}{\ensuremath{\vec{B}_{\rm POS}}}
\newcommand{\pmas}{\ensuremath{p_{\rm MAS}}}

\newcommand{\dlxx}{\ensuremath{{\cal D}_{\ell}^{XX}}}

\newcommand{\dlee}{\ensuremath{{\cal D}_{\ell}^{EE}}}
\newcommand{\dlbb}{\ensuremath{{\cal D}_{\ell}^{BB}}}
\newcommand{\dlte}{\ensuremath{{\cal D}_{\ell}^{TE}}}

\newcommand{\mlxx}{\ensuremath{{\cal M}_{\ell}^{XX}}}

\newcommand{\etob}{$E$-$B$}
\newcommand{\phase}{template}
\newcommand{\phases}{templates}

\begin{document}

\title{Dust polarization modelling at large scale over the northern Galactic cap using EBHIS and \planck\ data}
\author{Debabrata Adak\thanks{Corresponding author: debabrata@iucaa.in}  \inst{1}, Tuhin Ghosh\thanks{Corresponding author: tghosh@niser.ac.in} \inst{2}, Francois Boulanger \inst{3}, Urmas Haud \inst{4},  Peter Kalberla \inst{5},  Peter G. Martin  \inst{6},  \\ Andrea Bracco  \inst{3}, Tarun Souradeep  \inst{1,7}}

\institute{
Inter University Centre for Astronomy and Astrophysics, Post Bag 4, Ganeshkhind, Pune-411007, India \\
\and
School of Physical Sciences, National Institute of Science Education and Research, HBNI, Jatni 752050, Odisha, India \\
\and
Laboratoire de Physique de l'\'{E}cole Normale Sup\'{e}rieure, ENS, Universit\'{e} PSL, CNRS, Sorbonne Universit\'{e}, Universit\'{e} de Paris, F-75005 Paris, France \\ 
\and
Tartu Observatory, 61602 T\~{o}ravere, Tartumaa, Estonia\\
\and
Argelander-Institut f\"{u}r Astronomie, Universit\"{a}t Bonn, Auf dem H\"{u}gel 71, D-53121 Bonn, Germany\\
\and
CITA, University of Toronto, 60 St. George St., Toronto, ON M5S 3H8, Canada\\
\and
Indian Institute of Science Education and Research, Dr. Homi Bhabha Road, Ward No. 8, NCL Colony, Pashan, Pune-411008, Maharashtra, India
 }

\abstract{The primary source of systematic uncertainty in the quest for the $B$-mode polarization of the Cosmic Microwave Background (CMB) introduced by primordial gravitational waves is polarized thermal emission from Galactic dust. Therefore, accurate characterization and separation of the polarized thermal dust emission is an essential step in distinguishing such a faint CMB $B$-mode signal.  We provide a modelling framework to simulate polarized thermal dust emission based on the model described in Ghosh et al. (2017), making use of both the $\planck$ dust and Effelsberg-Bonn \hi\ surveys over the northern Galactic cap. Our seven-parameter dust model, incorporating both \hi\  gas in three different column density templates as a proxy for spatially variable dust intensity and a phenomenological model of Galactic magnetic field, is able to reproduce both 1- and 2-point statistics of the observed dust polarization maps seen by \planck\ at 353 GHz over a selected low-column density region in the northern Galactic cap. This work has important applications in assessing the accuracy of component separation methods and in quantifying the confidence level of separating polarized Galactic emission and the CMB $B$-mode signal, as is needed for ongoing and future CMB missions.}

\titlerunning{Modelling and simulation of polarized dust emission} 
\authorrunning{D. Adak et al.} 
\maketitle

\section{Introduction} \label{sec:intro}

Gravitational waves, possibly generated during the inflation epoch introduce $B$-mode polarization in the cosmic microwave background (CMB; \citealt{starobinski:1979,Fabbri:1983}). The measurement of this primordial CMB $B$-mode signal is the intense focus of ongoing and proposed CMB experiments. Galactic polarized foregrounds, especially polarized thermal dust emission \citep{M07,Bruce:2009,Vaillancourt:2012,planck2014-XXII,Ashton:2018,planck2016-l11B}, also produce a $B$-mode signal at microwave frequencies \citep{planck2014-XXX,planck2016-l11A}. Accurate modelling and separation of polarized dust emission is therefore a necessary step in the detection of a primordial CMB $B$-mode signal. Residual foregrounds due to imperfect component separation can be misinterpreted as detection of primordial CMB $B$-modes \citep{Remazeilles:2016}. 

Polarized thermal dust emission is expected from aspherical grains aligned with respect to the local Galactic magnetic field (GMF; \citealt{stein:1966}). Many distinct mechanisms have been elaborated to account for the alignment of dust grains, starting from the first quantitative approach by \cite{Davis&Greenstein:1951}; for a review, see \cite{A.Lazarian:2007} and \cite{Andersson:2015}. 

Empirically, the all-sky map of dust polarization from \planck\ at 353 GHz \citep{planck2014-a01} reveals a connection between the dust intensity structures and the local orientation of the GMF projected on the plane of the sky (\Bpos) \citep{planck2014-XXXII,planck2015-XXXV}. In the low-column-density or diffuse interstellar medium (ISM), the orientation of the dust intensity structures is observed to be preferentially parallel to \Bpos\ \citep{planck2014-XXXII}. Going to high-column-density regions, including star-forming molecular clouds, the preferred orientation of the dust intensity structures changes from parallel to perpendicular \citep{planck2015-XXXV,Soler17,Jow18}. This change in relative orientation was predicted using sub- and trans-Alfvenic magnetohydrodynamic (MHD) simulations  \citep{Soler13}, and highlights the role of the magnetic field during the formation of the molecular clouds.

Away from the Galactic plane, \Planck\ polarization maps reveal a large scatter in the distributions of the polarization fraction, $p$, and the polarization angle, $\psi$ \citep{planck2014-XIX}. A comparison with maps computed from a simulation of MHD turbulence in \cite{planck2014-XX} shows that the large scatter of $p$ is associated with variations in the orientation of the GMF along the line of sight (LOS), causing a depolarization effect. The simulation also reproduces the inverse relationship between $p$ and polarization angle dispersion function, $\mathcal{S}$ \citep{planck2014-XIX}. Subsequently, \cite{planck2016-XLIV} connected the distributions of $p$ and $\psi$ with the amplitude of turbulent magnetic field using a phenomenological model. This model is discussed further and compared with the \Planck\ data in \citet{planck2016-l11B}.

The \Planck\ maps were also used to measure power spectra of dust polarization. At  intermediate and high Galactic latitudes, in the multipole range, $40 < \ell < 600$, the ratio of dust $EE$ to $BB$ power ($\dlee/\dlbb$) is found to be around 2 \citep{planck2014-XXX,planck2016-l11A}. \cite{planck2015-XXXVIII} show that both the
\etob\ power asymmetry and the correlation between dust temperature and $E$ mode polarization, i.e. the dust $TE$ correlation, can be accounted for by the alignment between the orientation of the filamentary structure of interstellar matter in the diffuse ISM, as traced by total dust emission, and the orientation of the GMF, as inferred from dust polarization \citep{planck2014-XXXII,planck2015-XXXVIII}. \cite{Caldwell:2017} and \cite{Kandel:2018} investigated whether these observed properties could be related to turbulence in the magnetised ISM, considering the contributions from slow, fast, and Alfven MHD modes. \cite{ Caldwell:2017} conclude that the \etob\ power asymmetry and positive $TE$ correlation cannot both be accounted for with their model. Using the same theoretical framework, \cite{Kandel:2017} and \cite{Kandel:2018} reached the opposite conclusion, assuming that any correlation between the gas density and the magnetic field is negligible. However, this assumption is challenged by the above-mentioned observed alignment between the filamentary structure and the magnetic field. Filamentary structures identified in \hi\ 21-cm line spectroscopic data cubes, which were shown to trace density structure in the cold neutral medium (CNM) \citep{Clark19}, are also found to be aligned with the GMF \citep{clark:2014,clark:2015,Martin:2015,Kalberla:2016}. Furthermore, \citet{Clark18} reported a correlation between the polarization fraction, $p$, and the degree of coherence of the orientation of \hi\ emission features along the LOS, later used by \citet{Clark_Hensley19} to model dust polarization. Clearly, \hi\ data contain valuable information on the structure of the GMF in the neutral atomic ISM.  

\citealt{T_Ghosh:2017} (hereafter \tgmodel) present a phenomenological model of dust polarization. This \tgmodel\ model combines the framework introduced by \cite{planck2016-XLIV} with a decomposition of \hi\ emission data into three distinct maps of \hi\ column density referred to as `\hi\ templates'. By adjusting just a few parameters, the \tgmodel\ model reproduces the one- and two-point statistical properties of dust polarization over that fraction of the southern Galactic cap (defined by the region of $b \leq -30\deg$) where dust
and \hi\ emission are well correlated. 
In this paper, we use this framework to fit the \planck\ PR3 data \citep{planck2016-l01} over a fraction of the northern Galactic cap at $b \geq 30\deg$. The main goal is to extend the sky area available to fit and test the \tgmodel\ model. 

This paper is organised as follows. 
In Sect.~\ref{sec:data}, we describe the data used from \planck\ and the Effelsberg-Bonn \hi\ Survey (EBHIS).
Section~\ref{sec:3.0} describes the choice of sky region and \hi\ templates used in our analysis.
We summarise the statistical properties of the dust polarization over the selected region in Sect.~\ref{sec:4}. 
In Sect.~\ref{sect:3}, we briefly describe the phenomenological modelling framework. 
Section~\ref{sec:5} describes how the model parameters were evaluated.
We present separate aspects of the results in Sects.~\ref{sec:6} and \ref{sec:7}. 
Finally, we summarise our results in Sect.~\ref{sec:8}. 
In Appendix~\ref{sec:A1}, we explore a model based on alternative \hi\ templates.

\section{Datasets used} \label{sec:data}

\subsection{\planck\ dust polarization maps} \label{sec:2.1}

In this paper, we use publicly available \planck\ PR3 data\footnote{\url{http://www.cosmos.esa.int/web/planck/pla}}  \citep{planck2016-l01} at 353 GHz to study the statistical properties of the dust polarization. These maps are produced using only the polarization-sensitive bolometers (PSBs) and are expressed in thermodynamic temperature units (\kcmb, \citealt{planck2016-l03}). We also use various subsets of the \planck\ polarization data at 353 GHz, namely, the half-mission maps (HM1 and HM2), yearly surveys (Y1 and Y2), and odd and even surveys (O and E) to debias the effect of instrumental noise (Sect.~\ref{sec:4}).

For dust intensity, we use the generalised needlet internal linear combination (GNILC) processed Stokes $\textit{I}$ map \citep{planck2016-l04}. \planck\ does not have the ability to measure absolute emission, and so it is necessary to correct for the zero level of the dust intensity map. By construction, the GNILC dust intensity map has a cosmic infrared background (CIB) monopole contribution of  452\,\mukcmb\  at 353 GHz, which needs to be subtracted \citep{planck2016-l03}. To correct for the Galactic \hi\ offset and the contribution from dust emission associated with \hii\ emission \citep{planck2016-l11B}, a `fiducial' offset of 63\,\mukcmb\ is added back to the CIB-subtracted \GNILC\ dust intensity map (hereafter, we refer to this final map as \ignilc). The fiducial offset has an uncertainty of 40\,\mukcmb, with an associated `low' and `high' offset of 23 \mukcmb\ and 103 \mukcmb, respectively. 

The \planck\ Stokes $Q$ and $U$ maps at 353 GHz have a beam resolution of 4\farcm82 (FWHM), and the \ignilc\ map has the beam resolution of 5\arcm\ (FWHM). To increase the signal-to-noise ratio, we smooth the dust Stokes $I$, $Q$, and $U$ maps to a common resolution of 60\arcm\ or 80\arcm\ (FWHM) and reproject on the \healpix\ grid\footnote{{\tt https://healpix.jpl.nasa.gov}}~\citep{Gorski:2005} with $\Nside = 128$. To compare our dust model with the \planck\ data, we work with the 60\arcm\ (FWHM) smoothed maps. 
However, to study the inverse relationship between $\mathcal{S}$ and $p$ (Sect.~\ref{sec:7})
we work with 80\arcm\ (FWHM) smoothed \planck\ data.

\subsection{EBHIS and HI4PI \hi\ data} \label{sec:2.2}

We use \hi\ 21-cm line spectroscopic data from EBHIS\footnote{EBHIS data are available at \url{http://cdsarc.u-strasbg.fr/viz-bin/qcat?J/A+A/585/A41}} \citep{kerp_EBHIS:2011,EBHIS:2016}, which mapped the Milky Way gas in the northern Galactic sky with the 100\,m telescope at Effelsberg. The survey has an angular resolution of 10\farcm8 (FWHM), spectral resolution $\delta v = 1.44$~\kms (FWHM), and rms brightness temperature uncertainty of 90 mK. Velocities are with respect to the local standard of rest (LSR). The EBHIS data are projected on a \healpix\ grid with $\Nside = 1024$ (${\theta}_{\rm pix}$  = 3\farcm4). The EBHIS data also form the northern part of the all-sky product HI4PI \citep{HI4PI:2016}.

For optically thin emission, the total \hi\ column density (\NHI) can be obtained by integrating the brightness temperature ($T_{\rm b}$) over velocity channels 
\citep{Wilson:2009}:
\begin{align}
\NHI & =  1.82\times 10^{18} \times\int  T_b \, dv \,\,\,\text{cm}^{-2} \, . 
\label{eq:2.2.2}
\end{align}

\section{Exploiting the \hi\  data} \label{sec:3.0}

\subsection{Region selection} \label{sec:2.3sec:4}

We describe the procedure to select the low-column-density region in the northern Galactic cap in which \hi\ and dust emission are highly correlated, making \hi\ a good proxy for dust.

Ultraviolet observations of molecular hydrogen, \hmol\ \citep{savage:1977,Gillmon:2006} and early results from \planck\ \citep{planck2011-7.12} show that 
dust emission associated with gas in the form of  \hmol\ becomes significant
for sight lines where the total \NH\ exceeds $4 \times 10^{20}\, \cm2$ \citep{Arendt:1998}.
Therefore, we restrict ourselves to the sky region in which $N_{\hi,50}$ is below a threshold $3.8 \times 10^{20}\, \cm2$, where $N_{\hi,50}$ from Eq.~\eqref{eq:2.2.2} is evaluated over the restricted velocity interval $|v| \leq 50$\,\kms. We explored extending the range up to $80$\,\kms\ and because there is little additional gas in this range our results below from the modelling analysis (Sects.~\ref{sec:5} to \ref{sec:7}) are robust. We also get the same model parameters for regions restricted by  $N_{\hi,50}$ thresholds in the range $3.4 - 4.0 \times 10^{20}\, \cm2$.
The sky fraction increases beneficially with threshold, but not significantly beyond the adopted  $3.8 \times 10^{20}\, \cm2$.   However, the model parameters change significantly for thresholds above $4 \times 10^{20}\, \cm2$, presumably because of the increased and unaccounted molecular fraction.

\begin{figure}[h]
\includegraphics[width=8.8cm]{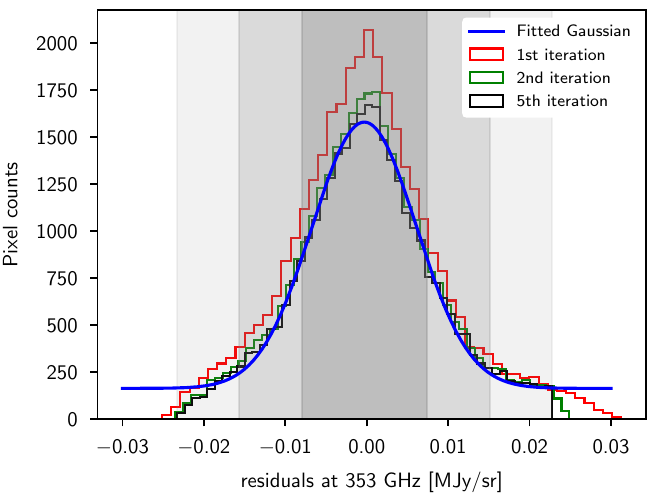}
\caption{Histogram of residuals at different iterations. The \textit{blue} curve is the Gaussian fitted to residuals at the final iteration. \textit{Gray} shaded regions are $1\,\sigma_{cg}$, $2\,\sigma_{cg}$, and $3\,\sigma_{cg}$ regions of the fitted Gaussian. }
\label{fig:resid}
\end{figure}

We build our mask using an iterative correlation analysis method as described in \cite{planck2013-XVII}. The initial mask contains unmasked pixels for which both the EBHIS and \planck\ data are available over the northern Galactic cap. At each iteration, we compute the linear correlation between \ignilc\ and total \NHI\ over unmasked pixels of the binary mask produced in the previous iteration. We compute residuals by subtracting the fit to the correlation from the observed dust emission for each unmasked pixel and find the standard deviation ($\sigma_{cg}$) characterising the Gaussian core of these residuals (Fig.~\ref{fig:resid}). 

\begin{figure}[h]
\includegraphics[width=8.8cm,height=8.8cm]{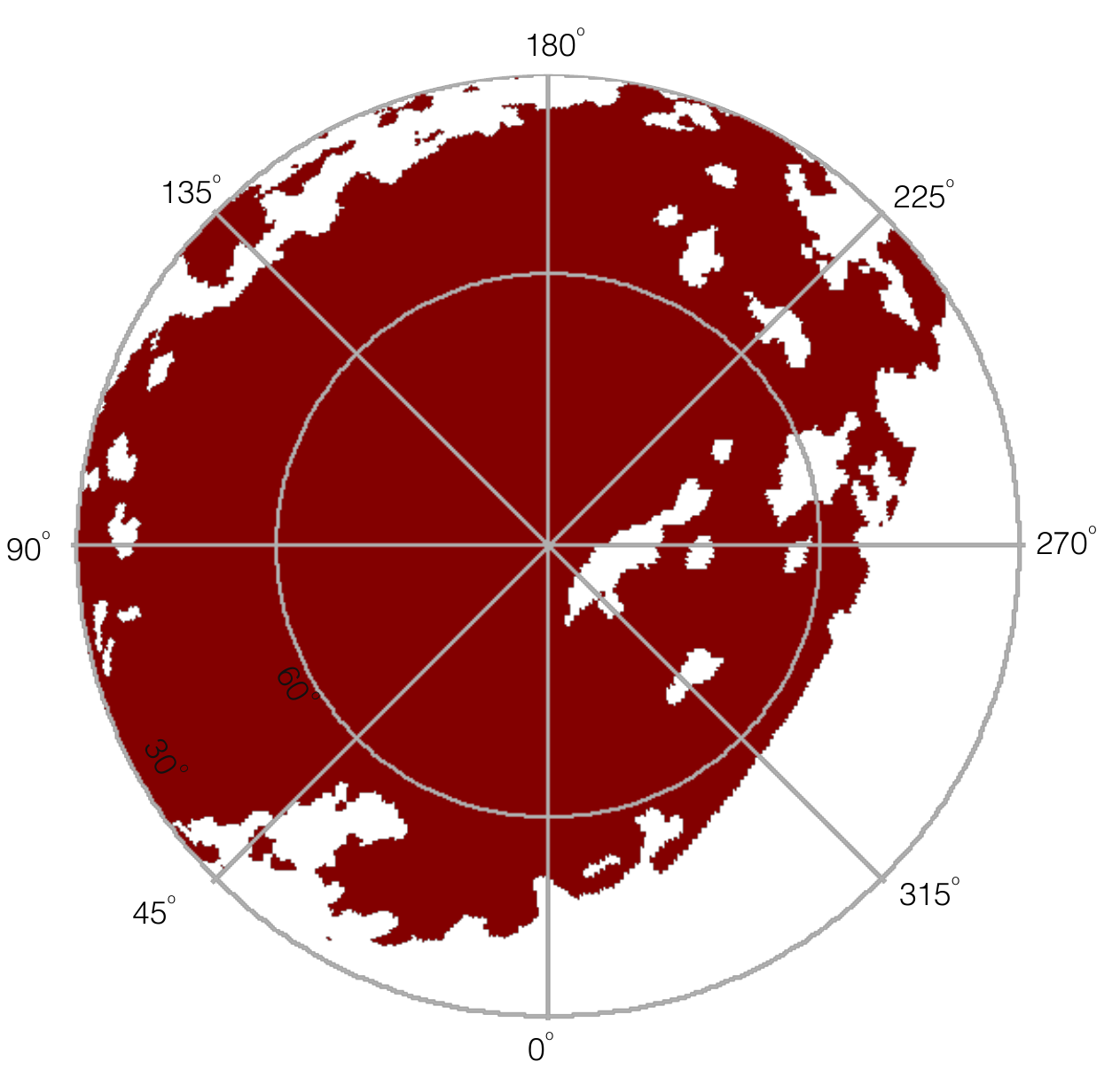} 
\caption{Binary mask defining \rgn, the $\textit{dark}$ regions selecting 65\,\% of the northern Galactic cap ($b \geq 30\deg$).}
\label{fig:mask}
\end{figure}

Next, the binary mask is updated by masking pixels for which the absolute value of the residual is greater than $3\,\sigma_{cg}$. A stable mask is obtained after five iterations, a region of $5900\,\text{deg}^2$ comprising 65\,\% of the northern Galactic cap ($b \geq 30\deg$), as  shown in Fig.~\ref{fig:mask}.   

Hereafter we refer to this region as \rgn.  
For later analysis, we apodize this binary mask by convolving it with a Gaussian window function of $2\deg$ (FWHM). After apodization, we have an effective sky fraction, $f_{\rm sky}^{\rm eff}$, of $0.143 $ (14.3\,\%). 
In the stable mask, there are some isolated islands with very few pixels. When we derive the full-sky angular power spectrum $C_{\ell}$ from the pseudo-spectrum, these isolated islands introduce correlations from very high $\ell$ to low $\ell$ via the mode-mode coupling matrix \citep{Hivon:2002}. In our analysis, we eliminate isolated islands containing less than 20 pixels using the \processmask\ routine of \healpix.

\begin{figure*}[hbt!]
\includegraphics[width=18cm]{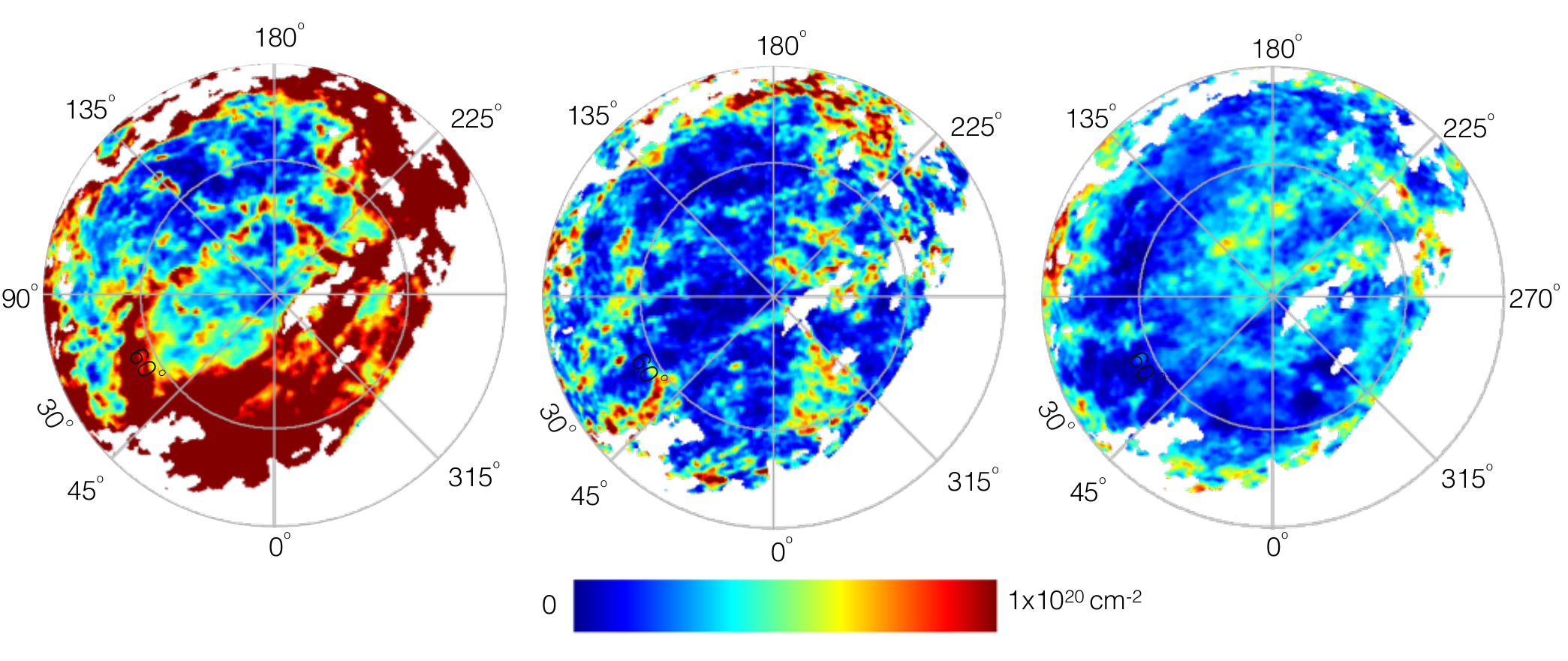} 
\caption{Northern Galactic orthographic projection of integrated \hi\ emission maps of the three \hi\ \phases\ called CNM (\textit{left panel}), UNM (\textit{middle panel}), and WNM (\textit{right panel}).  Over our selected region \rgn\ (Sect.~\ref{sec:2.3sec:4}), the mean column densities of the three \phases\ are $\left\langle N_{\hi}^{\rm c} \right\rangle  = 0.90\times 10^{20} \cm2$, $\left\langle N_{\hi}^{\rm u} \right\rangle  = 0.32 \times 10^{20} \cm2$, and $\left\langle N_{\hi}^{\rm w} \right\rangle  = 0.56 \times 10^{20} \cm2$, respectively. }
\label{figure:2.1}
\end{figure*}

Over \rgn\ the mean \hi\ column density over the restricted velocity interval is $\left\langle N_{\hi,50} \right\rangle  =1.85 \times 10^{20} \cm2$ and the mean dust intensity is $\left\langle \ignilc \right\rangle = 293\,\mukcmb$ at 353 GHz. The linear correlation between \ignilc\ and total \NHI\  yields a slope or emissivity $\epsilon_{353,50}$ of $137\,\mukcmb (10^ {20}\,\cm2)^{-1}$, where again the subscript 50 denotes the restricted velocity interval. This value is about 2\,\% higher than that found for the diffuse sky studied by \citet{planck2013-p06b}. The offset of $40\,\mukcmb$ is satisfactory, within the systematic uncertainty of the zero level (Sect.~\ref{sec:2.1}).

\subsection{\hi\ templates} \label{sec:phases}

The diffuse ISM is a complex turbulent multiphase medium. The \hi\ gas comprises two thermally stable phases, the cold and warm neutral medium (CNM and WNM, respectively), along with an additional thermally unstable phase at an intermediate temperature, hereafter referred to as the unstable neutral medium (UNM) following \tgmodel.  

\hi\ spectra can be used to build maps of \hi\ emission associated notionally with the CNM, UNM, and WNM. 
First, the observed brightness temperature profiles of \hi\ spectra are decomposed into Gaussian components \citep{Haud:2000,Haud:2013} under the assumption that the random velocities in any \hi\ `cloud' have a Gaussian distribution. Thus,
\begin{equation}
T_{\rm b}(v) = \sum_i T_{0}^i \, \exp\left[-\frac{1}{2} \left(\frac{v_{LSR} -v_{\rm c}^i}{\sigma^i} \right)^2 \right]  \, , 
\label{eq:2.2.1}
\end{equation}
where the summation is over all Gaussian components needed to describe the profile along a LOS, $v_{\rm LSR}$ 
is the velocity of the \hi\ gas, and $T_{0}^i$, $v_{\rm c}^i$, and ${\sigma}^i$ are the peak brightness temperature (expressed in Kelvin), central peak velocity, and 1\,$\sigma$ standard deviation of Gaussian component $i$, respectively. 

Following the detailed physically motivated prescription in section 2.2 of TG17, we produce CNM, UNM, and WNM column density maps. This involves using weight factors applied in summing the contributions from the Gaussian components. The weights involve two parameters of the polarization model of TG17 that constrain the range of centroid velocities of the Gaussians that contribute to each of the three maps. Here we simply adopt the TG17 values for these two parameters and the same weighting scheme, which should not be materially different in the north as compared to the south. TG17 noted the beneficial effect of having considerable column density in the three maps and that although the CMN contains some higher velocity dispersion gas, subdivision into two has little effect on the model. As we show below, this prescription does indeed lead to an acceptable polarization model for \rgn.
 
These three maps have an original resolution of 10\farcm8 (FWHM), which we have smoothed to a common resolution of 60\arcm (FWHM) at \Nside = 128. The resulting maps, to be used as \hi\ templates below, are shown in Fig.~\ref{figure:2.1}.
In Appendix.~\ref{sec:A1}, we present an analysis that uses alternative \hi\ templates based on a decomposition by \cite{Kalberla:2018}.

\section{Observed statistics from $\planck$ dust observations over the selected region} \label{sec:4}

In this section, we analyse the statistical properties of dust polarization over \rgn\ in both pixel and harmonic space.

\begin{figure*}[h]
\begin{multicols}{2}
    \includegraphics[width=\linewidth]{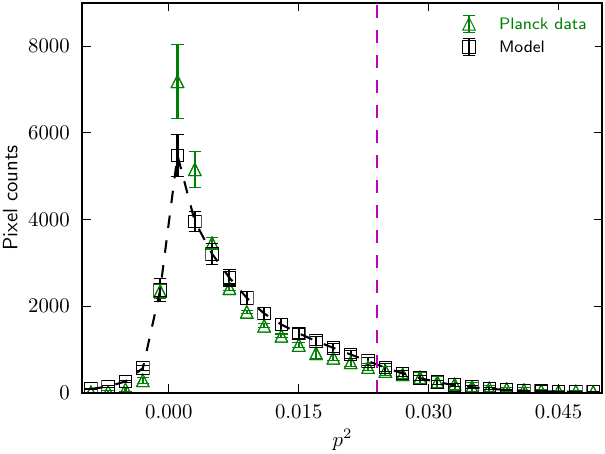}\par 
    \includegraphics[width=\linewidth]{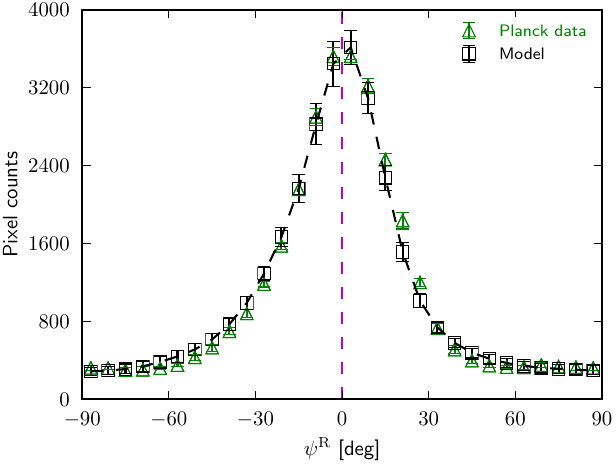}\par 
    \end{multicols}
\caption{\textit{Left panel:} Histogram of the square of the polarization fraction, $p^2$,  at 353\,GHz. \textit{Right panel:} Histogram of $\psi^{\rm R}$, the difference between the measured polarization angle at 353\,GHz and that computed for the mean GMF orientation over \rgn. $\textit{Green triangles}$ are mean $p_{\rm d}^2$ ($\psi_{\rm d}^{\rm R}$) estimated from \planck\ data subsets, and $ \textit{black squares}$ are mean $p_{\rm m}^2$ ($\psi_{\rm m}^{\rm R}$) estimated from 100 realisations of model maps. The vertical dashed line in the $\textit{left panel}$ corresponds to the model parameter $p_0 = 15.5$\,\% ($p_0^2 = 0.024$), while the vertical dashed line in the $\textit{right panel}$  corresponds to a model without a turbulent component of the GMF. }
\label{figure:5.1.1}
\end{figure*}

\subsection{Polarization fraction} \label{sec:4.1}

To avoid a bias from data noise, we compute the square of the polarization fraction ($p_{\rm d}^2$) combining independent subsets of the \planck\ data at 353\,GHz:
\begin{equation}
p_{\rm d}^2  = \left < \frac{Q_{\rm d}^{\rm s_1} \, Q_{\rm d}^{\rm s_2}  \, + U_{\rm d}^{\rm s_1} \, U_{\rm d}^{\rm s_2}}{\ignilc^2} \right > \, ,
\label{eq:5.1.1}
\end{equation}
where `d' stands for the \planck\ data, $(s_1, s_2)$ stands for different subsets of \planck\ data at 353\,GHz: $\big\{$(HM1, HM2); (Y1, Y2); (odd, even)$\big\}$, and $\left < ... \right >$ implies an average over different data subsets. In the left panel of Fig.~\ref{figure:5.1.1}, we present the histogram of $p_{\rm d}^2$. The  uncertainties (1\,$\sigma$) in $p^2_{\rm d}$ are the quadratic sum of data systematic errors estimated from the three data subsets and the uncertainty in \ignilc\ associated with the
determination of the zero level of Galactic emission. The mean values of $p_{\rm d}$ over \rgn\ are listed in Table~\ref{tab:1} for the different data subsets.

\begin{table}[h]
\begingroup
\newdimen\tblskip \tblskip=1pt
\caption{Mean polarization fraction at 353 GHz and orientation of the large-scale GMF using different subsets of the \planck\ data.}
\label{tab:1} 
\nointerlineskip
\vskip -3mm
\footnotesize
\setbox\tablebox=\vbox{
   \newdimen\digitwidth 
   \setbox0=\hbox{\rm 0} 
   \digitwidth=\wd0 
   \catcode`*=\active 
   \def*{\kern\digitwidth}
   \newdimen\signwidth 
   \setbox0=\hbox{+} 
   \signwidth=\wd0 
   \catcode`!=\active 
   \def!{\kern\signwidth}
\halign{
\hbox to .1cm{#\hfil}\tabskip 1.0em&
\hfil #\hfil &
\hfil #\hfil &
\hfil #\hfil &
\hfil #\hfil\tabskip=0pt\cr
\noalign{\doubleline}
 \omit Dataset  & \multispan{1} mean polarization  \hfil  &  subsets & \hfill large-scale GMF\cr 
  \omit  \hfil  &fraction ($p_{\rm d}$) & &  $l_0$ \hfill $b_0$   \cr
\noalign{\doubleline}
Half Missions  & 8.8\,\% &   HM1 & 61\deg\  \hfill  $-17\deg$ \cr
           					         & &     HM2 & 64\deg\ \hfill $-19\deg$ \cr    
\noalign{\vskip 4pt\hrule\vskip 6pt}
Years  &8.8\,\% &   YR1& 61\deg\ \hfill $-17\deg$  \cr
            &       &                YR2&  64\deg\ \hfill $-19\deg$ \cr
\noalign{\vskip 4pt\hrule\vskip 6pt}
Odd-Even  & 8.8\,\%  &   odd & 63\deg\ \hfill $-18\deg$ \cr
                  &               & even & 62\deg\ \hfill $-18\deg$ \cr
\noalign{\doubleline}
Mean & --\,  &   -- &   62\deg\  \hfill $-18\deg$ \cr
\noalign{\vskip 5pt\hrule\vskip 3pt}}}
\endPlancktablewide
\endgroup
\end{table}

\subsection{Polarization angle} \label{sec:4.2}

\begin{figure}[h]
\centering
\includegraphics[width=9.cm]{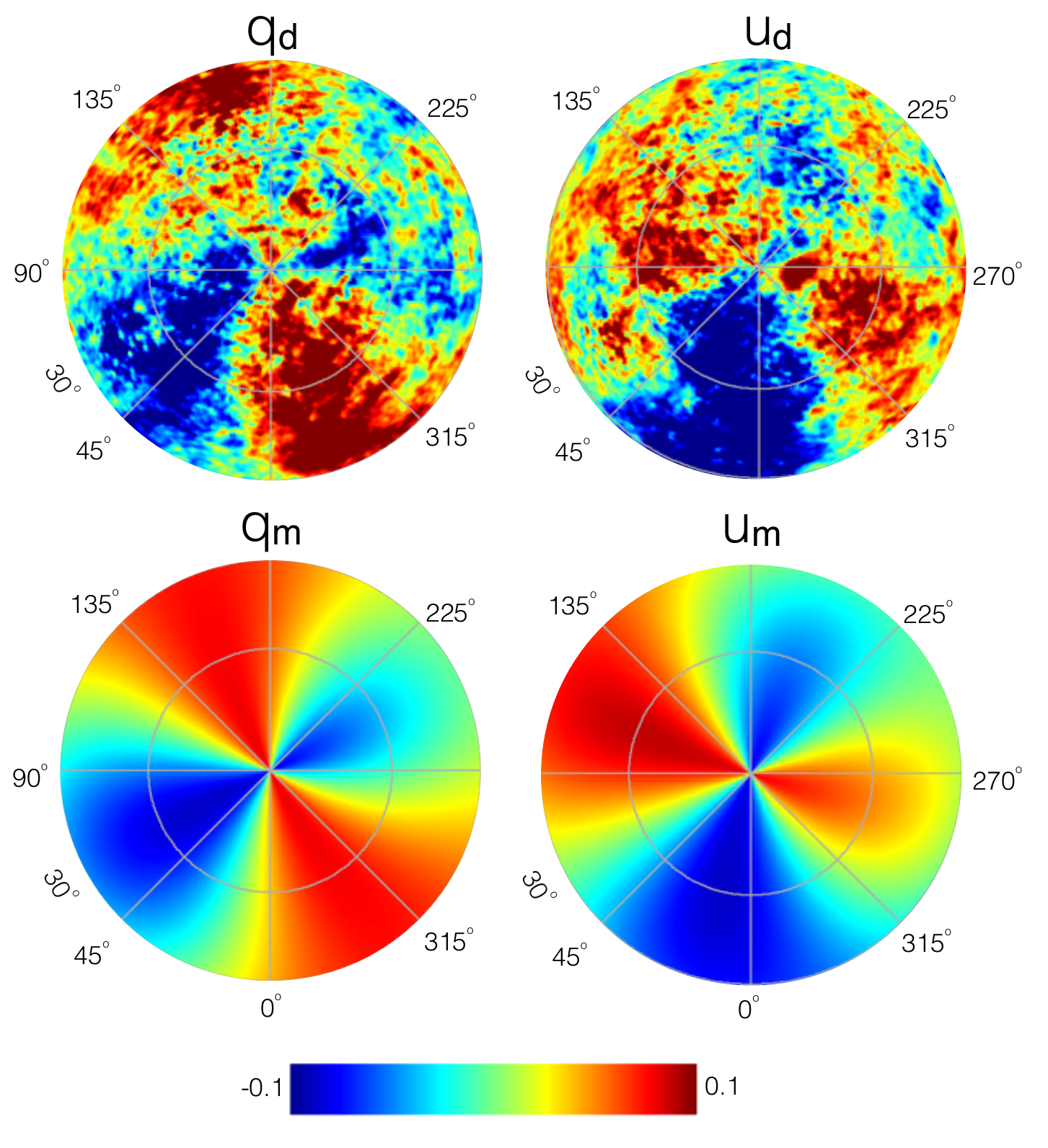} 
\caption{$\textit{Upper panel:}$ Northern Galactic orthographic projection of $q_{\rm d}$ (left) and $u_{\rm d}$ (right)  for the \planck\ data. 
$\textit{Lower panel:}$ Same for the best-fit model where the large-scale GMF is directed toward ($l_0, b_0$) = ($62\deg, -18\deg$) and scaled with a mean polarization fraction of 8.8\,\%.}
\label{fig:4.2.1}
\end{figure}

We computed the Stokes parameter ratios, $q_{\rm d} = Q_{\rm d}/\ignilc$ and $u_{\rm d}= U_{\rm d}/\ignilc$ from \planck\ data at 353\,GHz. As shown in the top panel of Fig.~\ref{fig:4.2.1}, maps of $q_{\rm d}$ and $u_{\rm d}$ centred at the northern Galactic pole reveal a `butterfly' pattern associated with an ordered large-scale GMF.

To determine $l_0$ and $b_0$ for this large-scale GMF, we fit these data with a normalised model of Stokes parameters 
as described in `step A' in \cite{planck2016-XLIV}.  
The best-fit values of $l_0$ and $b_0$ for different data subsets are noted in Table~\ref{tab:1} and the lower panel of Fig.~\ref{fig:4.2.1} presents the best-fit model.
These values, $(l_0, b_0) = (62\deg, -18\deg)$, give the `orientation' of the GMF, but not the direction; i.e. the $\pi$ ambiguity cannot be resolved using these data, though it can be resolved using rotation measures \citep{Xu19}. The corresponding values found in \tgmodel\ for the southern Galactic cap are $(l_0, b_0) = (74\deg, +24\deg)$. The change of sign of $b_0$ from north to south reflects the deformation of the large-scale GMF by the local bubble \citep{alves-et-al-2018}.


The mean best-fit values of $l_0$ and $b_0$ are then used to calculate a polarization angle map, ${\psi}_0$, for the mean GMF, where ${\psi}_0$ is the polarization angle of the model without the turbulent component of the GMF. 
At each sky pixel, we then rotate $Q$ and $U$ values 
with respect to the new reference angle ${\psi}_0$ at respective pixels using the relation,
\begin{align}
&\left[\begin{array}{c}
Q_{\rm d}^{\rm R}  \\
U_{\rm d}^{\rm R}  \end{array}\right] = 
\left(\begin{array}{cc}
\cos{2 \psi_0} & - \sin{2 \psi_0}   \\
\sin{2 \psi_0}  &  \cos{2 \psi_0}  \end{array}\right)
 \left[\begin{array}{c}
Q_{\rm d} \\
U_{\rm d}  \end{array}\right] \ ,
\label{eq:5.2.1}
\end{align}
where $Q_{\rm d}^{\rm R}$ and $U_{\rm d}^{\rm R}$ are Stokes parameters in the rotated frame. The polarization angle $\psi_{\rm d}^{\rm R}$  is derived using the relation 
$\psi_{\rm d}^{\rm R} = 0.5\, \mathrm{atan2}(- U_{\rm d}^{\rm R}, Q_{\rm d}^{\rm R})$,

where the two-argument function $\mathrm{atan2}(-U,Q)$ is used in place of $\mathrm{atan}(-U/Q)$ to avoid the $\pi$-ambiguity.
The minus sign is introduced to compute $\psi_{\rm d}^{\rm R}$ in the {\tt IAU} convention \citep{IAU:1996} from \healpix\ maps given in the \cosmo\  convention\footnote{\url{https://healpix.jpl.nasa.gov/html/intronode12.htm}} used by \planck.

\begin{figure*}[h!]
\begin{multicols}{2}
    \includegraphics[width=\linewidth]{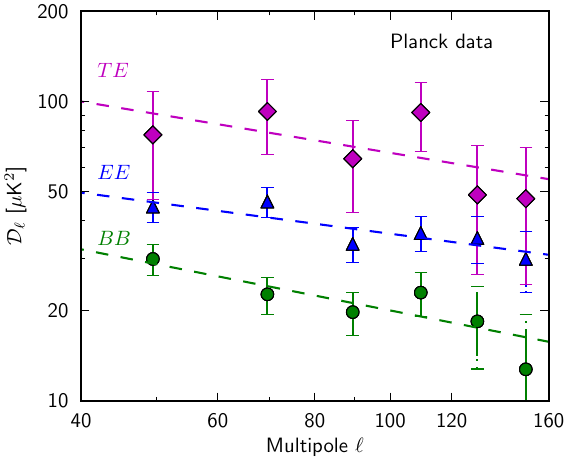}\par 
    \includegraphics[width=\linewidth]{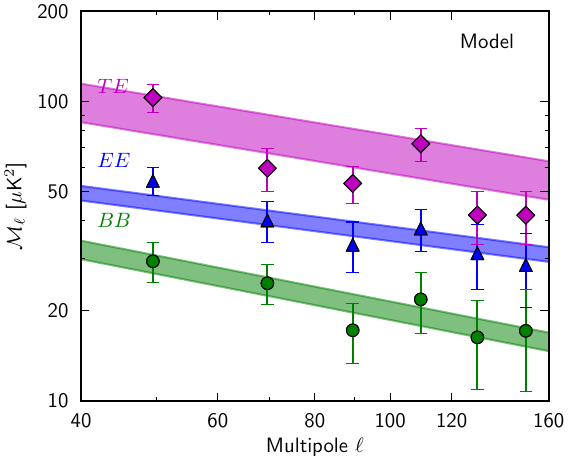}\par 
    \end{multicols}
\caption{$\textit{Left panel}$: Dust $EE,\, BB,$ and $TE$ cross-power spectra computed from the subsets of \planck\ data at 353 GHz over \rgn. $\textit{Dashed lines}$ represent 
the best-fit power laws. $\textit{Right panel}$: Similar plots as $\textit{left panel}$ computed from 100 realisations of the dust model maps. Error bars are 1\,$\sigma$ uncertainties as explained in the main text. The filled areas represent the \planck\ dust power spectra measurements over \rgn. }
\label{fig:5.3.1}
\end{figure*}

Figure~\ref{figure:5.1.1} (right panel) presents the distribution of $\psi_{\rm d}^{\rm R}$ over \rgn. The 1\,$\sigma$ error bars on the distribution of $\psi_{\rm d}^{\rm R}$ are computed from the standard deviation of the results from the six independent subsets of \planck\ data in Table~\ref{tab:1}. 
The 1\,$\sigma$ dispersion of $\psi_{\rm d}^{\rm R}$ derived from a Gaussian fit to the distribution is $16\pdeg6 \pm 0\pdeg2$.

\subsection{Dust power spectra} \label{sec:4.3}

We analyse the dust polarization power spectra at 353\,GHz over \rgn,\footnote{${\cal D}^{XX}_{\ell} \equiv \ell (\ell+1) C^{XX}_{\ell}/(2\pi$), where $XX\, = EE, BB, TE$.} \dlee , \dlbb, and \dlte, in the multipole range $40 \leq \ell \leq 160$.  These were computed using \xpol\ \citep{Tristram:2005} from cross spectra of three datasets: half-missions (HM1 $\times$ HM2), years (Y1 $\times$ Y2), and odd-even surveys (Odd $\times$Even). In Fig.~\ref{fig:5.3.1}, we present the mean of the three cross-power spectra in six multipole bins. The dust \dlee\ and \dlbb\ spectra are corrected for CMB contributions in harmonic space using the \planck\ best-fit power spectra \citep{planck2016-l06}, whereas we keep \dlte\ unaltered.
Uncertainties (1\,$\sigma$) on the binned dust power spectra are the quadratic sum of statistical noise computed by \xpol\ analytically, and systematic uncertainties computed from the standard deviation from the three \planck\ data subsets.

\begin{table}[h]
\begingroup
\newdimen\tblskip \tblskip=5pt
\caption{Fitted dust power spectra of the \planck\ data at 353\,GHz and of the dust model over \rgn.}
\label{table2}
\nointerlineskip
\vskip -3mm
\footnotesize
\setbox\tablebox=\vbox{
   \newdimen\digitwidth 
   \setbox0=\hbox{\rm 0} 
   \digitwidth=\wd0 
   \catcode`*=\active 
   \def*{\kern\digitwidth}
   \newdimen\signwidth 
   \setbox0=\hbox{+} 
   \signwidth=\wd0 
   \catcode`!=\active 
   \def!{\kern\signwidth}
\halign{
\hbox to 2.5cm{#\hfil}\tabskip 2.0em&
\hfil #\hfil &
\hfil #\hfil\tabskip=0pt\cr
\noalign{\doubleline}
 \omit Parameter \hfil & \planck\ 353\,GHz data & Dust model   \cr
\noalign{\doubleline}
$\alpha_{EE}$ & $-2.34\pm 0.16$ & $-2.56\pm 0.18$ \cr
$\alpha_{BB}$ & $-2.51\pm 0.21$ & $-2.56\pm 0.25$ \cr
$\alpha_{TE}$ & $-2.43\pm 0.38$ & $-2.75\pm 0.16$ \cr
\noalign{\doubleline}

$\chi^2_{EE} (N_{\rm d.o.f.} = 4)$ & 2.2 & 1.3 \cr
$\chi^2_{BB} (N_{\rm d.o.f.} = 4)$ & 1.8 & 1.3 \cr
$\chi^2_{TE} (N_{\rm d.o.f.} = 4)$ & 2.3 & 8.3 \cr
\noalign{\vskip 4pt\hrule\vskip 6pt}
$A_{EE}$ [$\mukcmbsq$] &  $39.1 \pm 2.2\phantom{0}$   & $39.9 \pm 2.7$ \cr
$A_{BB}$ [$\mukcmbsq$] &  $22.5 \pm 1.6\phantom{0}$   & $22.0 \pm 1.4$ \cr
$A_{TE}$ [$\mukcmbsq$] &  $74.2 \pm 10.9$ & $65.6\pm 4.0$ \cr
\noalign{\vskip 4pt\hrule\vskip 6pt}
$\langle A_{BB} /A_{EE} \rangle$ &$0.58 \pm 0.05$ & $0.57\pm 0.06$ \cr
$\langle A_{TE} /A_{EE} \rangle$ &$1.90 \pm  0.30$ &$1.85\pm 0.17$ \cr
\noalign{\vskip 5pt\hrule\vskip 3pt}}}
\endPlancktablewide
\par
\endgroup
\end{table}

We also checked the CMB correction at the map level. We subtracted the half-mission and odd-even component-separated CMB \smica\ and \sevem\ maps \citep{planck2016-l04} from the total \planck\ maps and then computed cross-power spectra using these four subsets of the maps. Subtraction of\, CMB does not introduce any noticeable changes in the dust polarization power spectra. 

Binned dust power spectra are well described by a power-law model, $\dlxx = A_{XX} (\ell/80)^{\alpha_{XX}+2}$, where $A_{XX}$ is the best-fit amplitude at $\ell =80$, $\alpha_{XX}$ is the best-fit spectral index, and again $XX=\{EE, BB, TE\}$ \citep{planck2014-XXX}. The best-fit values of $A_{XX}$ and $\alpha_{XX}$ and the respective values of $\chi^2$ are listed in the middle column of Table~\ref{table2}. 

The ratio of $BB$ to $EE$ power is about 0.6 over \rgn, consistent with the \etob\ asymmetry result of \cite{planck2014-XXX}. We detect a significant positive $TE$ correlation over \rgn, as in \tgmodel\ for the southern Galactic cap.

\section{Multiphase model of polarized dust emission} \label{sect:3}
We use the modelling framework from \tgmodel, which is based on a decomposition of \hi\ emission-line data and incorporates the phenomenological magnetic field model described in \cite{planck-XLIV:2016}.  We briefly describe the salient concepts.

We use the fact that dust emission and \NHI\ are correlated so that templates of \NHI\ can be used as proxies for spatially variable dust intensity.
For optically thin dust emission,  
the model Stokes parameters  $I_{\rm m}$, $Q_{\rm m}$, and $U_{\rm m}$ at 353 GHz can be written as 
\begin{align}
I_{\rm m} (\hat{\vec{n}}) &=  ~\sum_{i=1}^{N}  \left [1- p_0 \left ( \cos^2 \gamma^i(\hat{\vec{n}}) - \frac{2}{3} \right ) \right ]  \,  \epsilon_{353}  \,  \NHIi (\hat{\vec{n}})   \nonumber \\
Q_{\rm m}(\hat{\vec{n}}) &= ~ \sum_{i=1}^{N}    p_0  \cos^2 \gamma^i(\hat{\vec{n}})  \cos 2\psi^i(\hat{\vec{n}}) \,  \epsilon_{353}  \,  \NHIi (\hat{\vec{n}})  \label{eq:4.2}  \\
U_{\rm m}(\hat{\vec{n}}) &= - \sum_{i=1}^{N}   p_0  \cos^2 \gamma^i(\hat{\vec{n}})  \sin 2\psi^i(\hat{\vec{n}})  \,  \epsilon_{353} \,  \NHIi (\hat{\vec{n}}) \nonumber \, ,
\end{align}
where $N$ is the number of distinct \phases\ , $\hat{\vec{n}}$ is the direction vector, $p_0$ is a parameter related to the dust grain properties \citep{Lee:1985,Draine:2009,planck2014-XX}, $\gamma$ is the angle made by the local magnetic field with the plane of the sky, $\psi$ is the polarization angle measured counterclockwise from Galactic north,\footnote{{\tt IAU} convention, the minus sign producing  $U$ in the \cosmo\ convention.} 
$\epsilon_{353}$ is the dust emissivity at 353\,GHz for each \hi\ \phase, and \NHI\ is the column density within the particular \phase. 
In this work, $N=3$, which is represented by the three distinct templates for CNM, UNM, and WNM (Sect.~\ref{sec:phases}). 
This is a remarkably small number to describe the ISM. However, our focus is on demonstrating what essentials can  nevertheless be captured by a polarization model that is inherently minimalist and not dependent on fine-tuning.
For simplicity, $\epsilon_{353}$ is assumed to be the same for all three \phases. 

The connection between the dust polarization and the structure of the GMF via the angles $\gamma$ and $\psi$ is developed as in \cite{planck-XLIV:2016}. 
Following \citet{Jaffe:2010}, the GMF, \BT, is expressed as a vector sum of an ordered component (\Bo) and a fluctuating component (\Bt). Because most of the dust emission comes from the thin Galactic disk, \cite{planck2016-XLIV} assume that there is an ordered large-scale magnetic field, \Bo, in the solar neighbourhood (approximately 200 parsec). 
 The butterfly pattern discussed in Sect.~\ref{sec:4.2} supports the simplifying assumption that each \phase\ has the same $\Bo$, oriented toward the $l_0$  and $b_0$ that are already determined. However, the components of \Bt\ are different in each \phase\ and are taken from independent realisations of a Gaussian random field that has a power spectrum with $C_{\ell} \propto  \ell^{\alpha_{\rm M}}$ for $\ell \ge 2$.
The strength of $ |\Bt|$ relative to $|\Bo |$ is parameterised by $f_{\rm M}^i$.
This phenomenological model captures schematically the association 
between the structure of the multiphase ISM and that of the GMF, an interdependence that we consider to be essential for modelling the dust polarization.  This interdependence is usually ignored in 3D models of the GMF, for example those with Gaussian random magnetic fields as in \cite{Levrier18} and \cite{Wang20}. 
 However, our approach does not allow us to include the divergence-free constraint because it is not a 3D model.

Here, this key interdependence is introduced explicitly by aligning the model GMF with the structures in the \hi\ emission maps (see section 4.3 in \tgmodel), underlying the potential for $TE$ correlation and \etob\ power asymmetry.

We can use the proxy  $\NHI^i$ \phases\  as pure $E$-mode polarization maps, transforming them
to $Q$ and $U$ maps and then use these latter to compute maps of the polarization angle describing the orientation. 
 Given the evidence (Sect.~\ref{sec:intro}),
 as in \tgmodel ,\, we assume  alignment between the local \Bpos\ and dust structures represented by the CNM template and model the polarization angles $\psi^{\rm c}$ as above.
However, for the UNM and WNM \phases\ we assume no such alignment\footnote{In this case, there is little direct evidence for or against alignment, and again this is a minimalist approach. In the actual turbulent ISM the evolutionary relationship between gas and fields in thermal phases is complex, whether time-dependent or steady-state, and probably varies with the position because of its particular history and the ambient pressure, radiative environment, etc.} 
and so we apply a different procedure in which the polarization angles $\psi^{\rm u}$ and $\psi^{\rm w}$ follow from random Gaussian realisations of the components of \Bt\ \citep{planck2016-XLIV}. 
 We simulate the $\cos^2\,\gamma$ factor in Eqs.~\eqref{eq:4.2} along each LOS for each of three \phases\ using the relation \citep{planck2016-XLIV}
\begin{equation}
\cos^2\,\gamma =  1 - | \hat{\vec{B}} \, . \, \hat{\vec{r}}|^2  \ , 
\end{equation}
where $\hat{\vec{B}}$ is the unit vector of total GMF and $\hat{\vec{r}}$ is the unit vector for a given LOS.

In summary, the seven parameters in the dust model are as follows:
\begin{itemize}

\item  $\epsilon_{353}$ is the mean dust emissivity at 353\,GHz (all model Stokes parameters scale with this value). 

\item $p_0$ is a polarization parameter that combines the polarization degree of interstellar dust grains and their alignment efficiency. 

\item Galactic coordinates $l_0$ and $b_0$ determine the direction of the ordered component of the GMF. 

\item $f_{\rm M}^{\rm c}$ parameterises the relative strength of $|\Bt|$ and $|\Bo|$ in the CNM \phase. The corresponding parameter is taken to be the same in the UNM and WNM \phases\ and is denoted $f_{\rm M}^{\rm u/w}$. 

\item $\alpha_{\rm M}$ is the exponent of the power spectrum of the turbulent component of the GMF, assumed to be the same in each \phase.  

\end{itemize}

We simulate a set of 100 Monte-Carlo model realisations of Stokes $I_{\rm m}$, $Q_{\rm m}$, and $U_{\rm m}$ maps at 353\,GHz for a set of parameters. Because the \planck\ data have noise, we add two independent half-mission end-to-end noise realisations provided by \planck\ \citep{planck2016-l03} to each noise-free dust model polarization map to produce two maps ($Q$ and $U$) with independent noise. We do not add noise to the $I_{\rm m}$ maps because the \ignilc\ map that we are using has negligible noise at 353\,GHz.

\section{Method of constraining model parameters} \label{sec:5}

The best-fit orientation of \Bo\ is $l_0 = 62\deg$ and $b_0 = -18\deg$ (see Table~\ref{tab:1} in Sect.~\ref{sec:4.2}). Here, we describe how we fit the other five parameters of the model.

\subsection{Turbulent magnetic field in the CNM template} \label{sec:5.1}

From the $\NHI^{\rm c}$ \phase\, we compute the polarization angles using the method described in Sect.~\ref{sect:3}. In particular, we compute maps of $\cos 2\psi^{\rm c}$. In the CNM the assumed alignment of the local GMF with respect to the gas structures constrains the pair of parameters, $f_{\rm M}^{\rm c}$ and $\alpha_{\rm M}$, that characterise \Bt.

\begin{figure}[h]
\includegraphics[width=9.0cm]{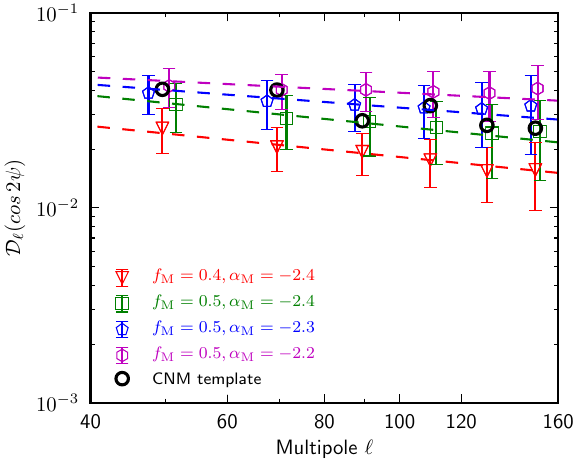}
\caption{Means of power spectrum amplitudes from simulated model maps of $\cos2 \psi$ within the multipole range $40 \leq \ell \leq 160$ for different values of $f_{\rm M}$ and $\alpha_{\rm M}$ and the best-fit power laws ($\textit{dashed lines}$). The error bars (1\,$\sigma$) are standard deviations computed from 100 realisations. $\textit{Black circles}$ are power spectrum amplitudes from the map of $\cos2 \psi^{\rm c}$ computed from the CNM template.}
\label{fig:7.1.1}
\end{figure}

To evaluate these constraints, following \tgmodel\ we simulate $\cos2 \psi$ for different values of the pairs assuming the adopted orientation of \Bo, 
using algebra described in section~4.1 of \cite{planck2016-XLIV}.  For each pair, we compute means of power spectrum amplitudes of $\cos2 \psi$ from 100 Monte-Carlo realisations over \rgn\ and fit these with a power-law model ${\cal D}_{\ell} \propto \ell^{\alpha_{\rm M} + 2}$ within multipole range $40 \leq \ell \leq 160$ (Fig.~\ref{fig:7.1.1}). Comparing the power spectra of the $\cos 2\psi^{\rm c}$ map with the simulated spectra of $\cos2 \psi$ for different pairs of $f_{\rm M}$ and $\alpha_{\rm M}$, we adopt the following pair of parameter values for the CNM \phase: $f_{\rm M}^{\rm c} = 0.5^{+0.05}_{-0.04}$ and $\alpha_{\rm M} = -2.4 \pm 0.1$.

Applying the same procedure to the UNM and WNM \phases\ results in a value of $f_{\rm M}^{\rm u/w}$ much higher than $f_{\rm M}^{\rm c}$ for the CNM \phase. However, as seen next, the model fit does not support such high turbulence in the UNM or WNM. As discussed in Sect.~\ref{sec:5.2}, the model is in favour of a very low turbulence in both the UNM and WNM, with $f_{\rm M}^{\rm u/w} = 0.1$.

\subsection{Additional model parameters}\label{sec:5.2}

We determine the three remaining model parameters, $\epsilon_{353}$, $p_0$, and $f_{\rm M}^{\rm u/w}$, by jointly minimising the following two expressions for $\chi^2$. One is in harmonic space (Fig.~\ref{fig:5.3.1}, Table~\ref{table2}):
\begin{equation}
\chi_{XX}^2 = \sum_{\ell_{\rm min}}^{\ell_{\rm max}} \left[ \frac{\dlxx - \mlxx ( \epsilon_{353}, p_0, f_{\rm M}^{\rm u/w})}{\sigma_{\ell}^{XX}} \right]^2 \, ,
\label{eq:chipower}
\end{equation}
where $XX=\{EE, BB, TE\}$, \dlxx\ and \mlxx\ are binned power spectra of dust within the multipole range $40 \leq \ell \leq 160$ for the data and model, respectively, and $\sigma_{\ell}^{XX}$ are the corresponding standard deviations.  
 The other is in pixel space (Fig.~\ref{fig:7.1.2}):
\begin{equation}
\chi^2 = \sum_{i=1}^{N_{\rm pix}} \left[ \ignilc - s I_{\rm m} (\epsilon_{353}, p_0, f_{\rm M}^{\rm u/w}) - o \right]^2 \, .
\label{eq:ioffset}
\end{equation}
Through combining statistics of the polarized dust observations in both harmonic and pixel space, we break the degeneracy between the two model parameters $\epsilon_{353}$ and $p_0$ in Eq.~\eqref{eq:4.2}. 
The $\chi^2$ of the T$-$T correlation between \ignilc\ and $I_{\rm m}$ is minimised in pixel space over \rgn. To match the observed dust amplitude at 353 GHz, the value of $s$ should be close to 1 and is kept so by adjusting $\epsilon_{353}$ during the iterative solution of the two equations.

The best-fit values are $\epsilon_{353} = 146\,\mukcmb (10^{20}\,\cm2)^{-1}$, $p_0 = 15.5$\,\%, and $f_{\rm M}^{\rm u/w} = 0.1$. Compared to the values of $\epsilon_{353}$ and $p_0$ found by \tgmodel\ in their analysis of the southern Galactic cap, our values differ by factors of 1.2 and 0.84, respectively.

\section{Comparison of the model with \planck\ observations} \label{sec:6}

Using the methods detailed in Sect.~\ref{sec:5}, we determined parameters of the Stokes dust model such that several statistical properties computed from the model match the \planck\ dust observations at 353 GHz over \rgn\ in both pixel and harmonic space. 
The quality of these matches to constraining the \Planck\ polarization data is demonstrated below. Furthermore, we show in Sect.~\ref{sec:7} that the dust model is able to reproduce the observed inverse relationship between the polarization angle dispersion and the polarization fraction over \rgn, even though these statistics were not used in determining the model  parameters.

\subsection{Polarization fraction} \label{sec:6.1}

We compute the mean square of the polarization fraction, $p_{\rm m}^2$, from 100 realisations using Eq.~\eqref{eq:4.2}. The error bars on the mean histogram are the standard deviations estimated over the 100 realisations. As shown in Fig.~\ref{figure:5.1.1} (left), our model reproduces the statistical distribution of $p_{\rm d}^2$ quite accurately, including the negative values of $p_d^2$ and the largest values beyond $p_0^2$.

\subsection{Polarization angle} \label{sec:6.2}

We compute the polarization angle, $\psi^{\rm R}_{\rm m}$, from 100 realisations of the model, following the method described in Sect.~\ref{sec:4.2} and used for the data.
For each sky realisation $i$, we fit the ordered component of GMF using $l_0^i$ and $b_0^i$ and from the map of the mean polarization angle of the GMF $\psi_0^i$.  At each sky pixel in the realisation of the Stokes model maps we then rotate $Q_i$ and $U_i$ with respect to the new reference angle $\psi_0^i$ at respective pixels using Eq.~\eqref{eq:5.2.1}. The polarization angle $\psi^{\rm R}_i$ is then derived from the rotated $Q_i$ and $U_i$ maps. The mean polarization angle, $\psi^{\rm R}_{\rm m}$, for the model is computed by taking the mean of the histograms of $\psi^{\rm R}_i$ distributions for each bin. Similarly, the uncertainties for each bin of the histogram are computed from the standard deviation of 100 realisations. As shown in Fig.~\ref{figure:5.1.1} (right), our model reproduces the statistical distribution of $\psi^{\rm R}_{\rm d}$. The dispersion of $\psi^{\rm R}_{\rm m}$ derived from a Gaussian fit is $17\pdeg5 \pm 0\pdeg4$, which is slightly higher than that for the data $\psi^{\rm R}_{\rm d}$: $16\pdeg6 \pm 0\pdeg2$. The difference might come from the fact that the model is not fitted to the \planck\ data at $\ell < 40$, which cannot be accomplished accurately over the limited area of \rgn.

\subsection{Dust polarization spectra} \label{sec:6.3}

We compute mean cross-power spectra from 100 sets of two independent realisations of the Stokes model maps. We fit the binned power spectra with power laws:  $\mlxx = A_{XX} (\ell/80)^{\alpha_{XX}+2}$ within the multipole range $40 \leq \ell \leq 160$. Error bars in each bin are the standard deviations computed from 100 realisations. Best-fit amplitudes, $A_{XX}$ and exponents, $\alpha_{XX}$ along with the corresponding $\chi^2$ values, are listed in the right column of Table~\ref{table2}. We compare the results with the \planck\ dust power spectra in Fig.~\ref{fig:5.3.1}. The \dlee\ and \dlbb\ amplitudes and their ratio match the values from the data accurately. The \dlte\ amplitude for the dust model is slightly lower, but consistent with the \planck\ data within its uncertainty.

\begin{figure}[h]
\includegraphics[width=9.0cm]{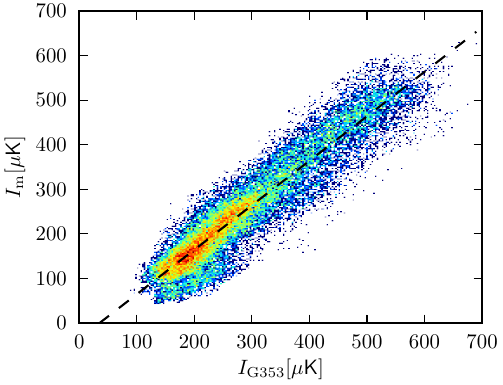}
\caption{Correlation plot between $I_{\rm m}$ and \ignilc\ at 353\,GHz. The $\textit{black dashed}$ line is the result of the joint optimisation of Eqs.~\eqref{eq:chipower} and \eqref{eq:ioffset}.
}
\label{fig:7.1.2}
\end{figure}

\subsection{Total intensity} \label{sec:6.4}

Figure~\ref{fig:7.1.2} shows the tight correlation between $I_{\rm m}$ and \ignilc\ at 353\,GHz, which demonstrates that statistically, the model dust intensity reproduces the observed \Planck\ dust intensity. The best-fit dashed line shown has slope $s = 1.0$ and offset $o = -35\,\mukcmb$, again within the uncertainty of the zero level in the intensity, $40\,\mukcmb$ (Sect.~\ref{sec:2.1}).

\begin{figure}[h!]
   \includegraphics[width=\linewidth,trim=0 -20 0 0, clip=true]{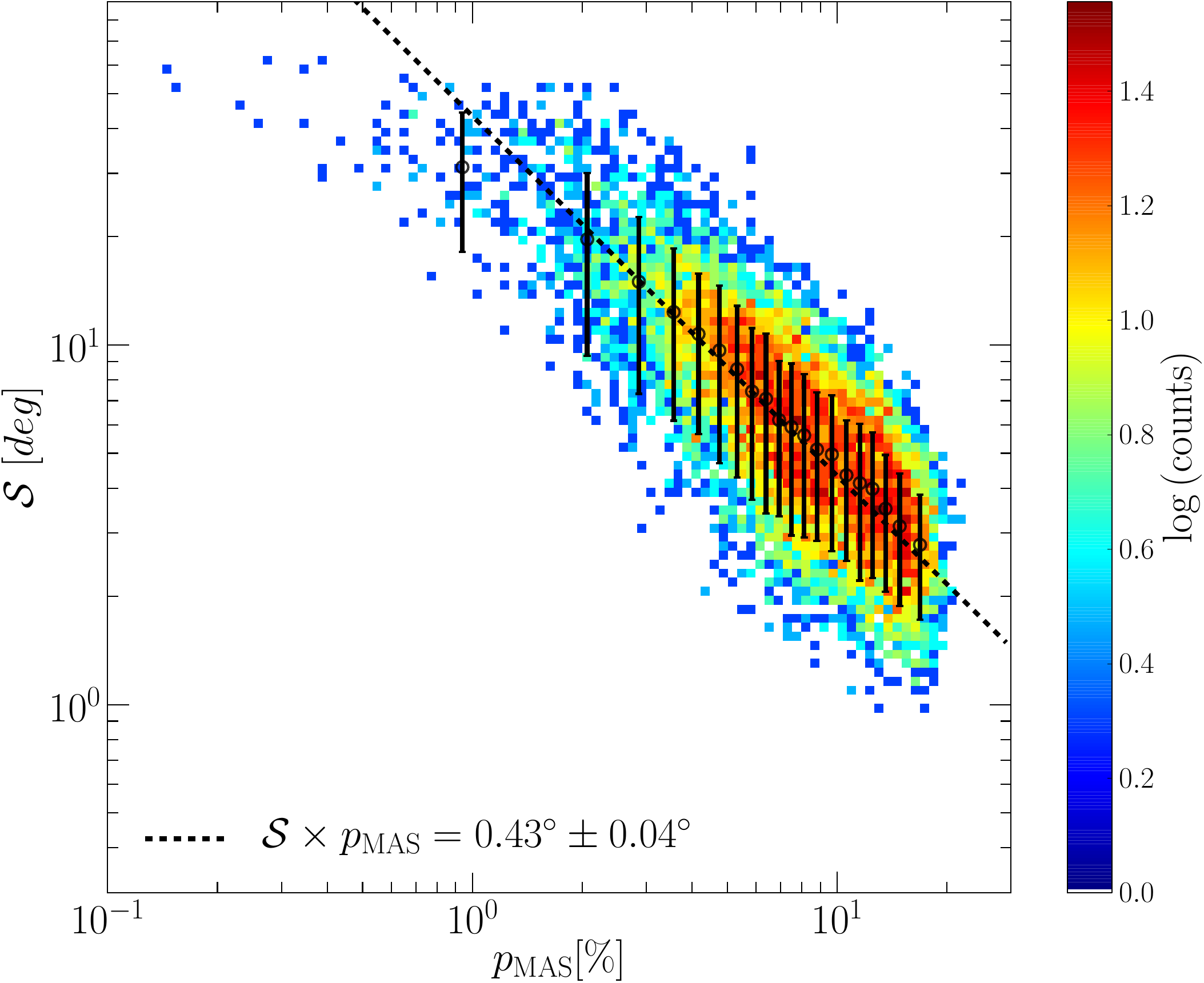}\par 
   \includegraphics[width=\linewidth]{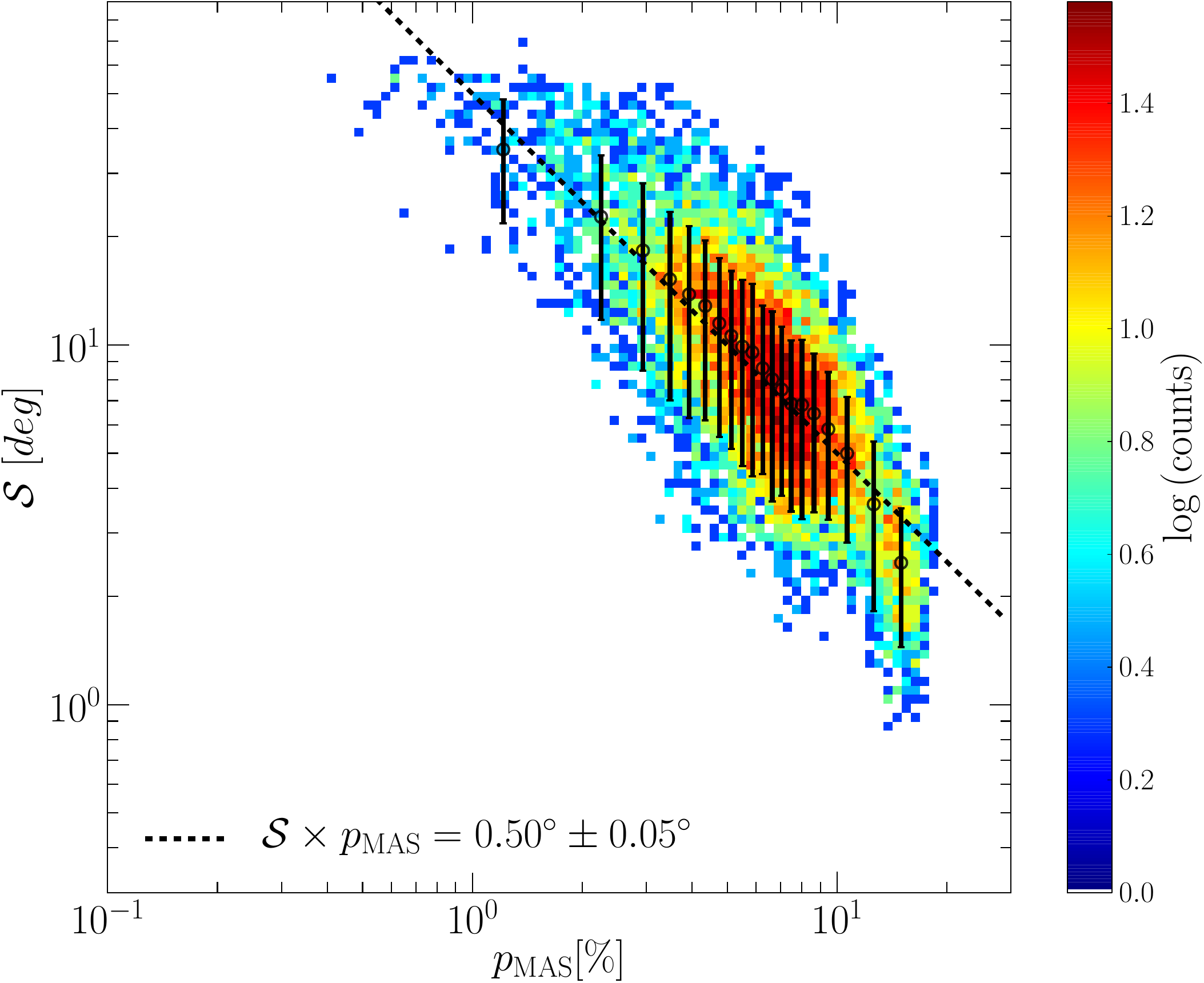}\par 
\caption{$\textit{Top panel}$: Two-dimensional histogram of the joint distribution of $\mathcal{S}$ and \pmas\ over \rgn\ for the \planck\ data at a resolution of 80\arcm\ with lag $\delta = 40\arcm$. 
The black circles show the mean of $\mathcal{S}$ in bins of \pmas\ containing the same number of map pixels. Error bars are the standard deviation of $\mathcal{S}$ in each bin. The 
black dotted line is a fit of the running mean. 
$\textit{Bottom panel}$: Same as $\textit{top panel}$, but for the model.}
\label{figure:s-p anti-correlation.1}
\end{figure}

\section{Inverse relationship between the polarization angle dispersion and the polarization fraction} \label{sec:7}

The polarization angle dispersion function, $\mathcal{S}$, was introduced in \cite{planck2014-XIX} to quantify the local dispersion of the dust polarization angle for a given angular resolution and lag $\delta$.
An inverse relationship between $\mathcal{S}$ and \pmas\ was found, where \pmas\ is the modified asymptotic estimator of the polarization fraction \citep{Plaszczynski:2014}.
That analysis was done over 42\,\% of the sky at low and intermediate Galactic latitudes at a resolution of 1\deg\ (FWHM) with a lag $\delta = 30\arcm$. 
\cite{planck2016-l11B} extended this work over the full sky using \GNILC\ processed intensity and polarization maps at a resolution of 160\arcm\ (FWHM) with $\delta = 80\arcm$ and found that $\mathcal{S} \propto \pmas^{-1}$.

Here, we study the relationship between $\mathcal{S}$ and \pmas\ over \rgn\ using 353 GHz maps at a resolution of 80\arcm\ (FWHM) with $\delta = 40\arcm$. The noise bias parameter needed for \pmas\ \citep{Montier-I:2015,Montier-II2015} is estimated from the smoothed noise covariance matrices, ${\sigma}_{II}, {\sigma}_{IQ}, {\sigma}_{IU}, {\sigma}_{QQ}, {\sigma}_{QU}, {\sigma}_{UU}$ at a resolution of 80\arcm\ (FWHM) and $\Nside = 128$.  

In the \textit{top panel} of Fig.~\ref{figure:s-p anti-correlation.1}, we present the two-dimensional joint distribution of $\mathcal{S}$ and \pmas\ for the \planck\ data over \rgn. The running mean of $\mathcal{S}$ in each bin of \pmas\ follows an inverse relationship,  $\mathcal{S} = (0\pdeg43 \pm 0\pdeg04 )/p_{\rm MAS}$. 
In the $\textit{bottom panel}$ of Fig.~\ref{figure:s-p anti-correlation.1}, we show the corresponding $\mathcal{S}$ - \pmas\ distribution for our model. This too has an inverse relationship and the slope $0\pdeg50 \pm 0\pdeg05$ is very close to that for the \planck\ data.  This is remarkable given that this statistical property of dust polarization has not been exploited in the fit of our model parameters.

\section{Summary and Discussion} \label{sec:8}

In this paper, we analyse the statistical properties of the \planck\ dust polarization maps at 353 GHz at a 60\arcm\ resolution (FWHM) over a low-column-density region that accounts for 65\,\% of the northern Galactic cap at latitudes larger than $30^\circ$ (\rgn).
We make use of the multiphase dust polarization model described in \tgmodel. The model shows how the dust polarization across the sky can be approximated on the basis of proxy \hi\ data: in particular, three independent \phases\ notionally representing the contributions from the CNM, UNM, and WNM.
The model has seven adjustable parameters whose values are determined by reproducing \planck\ observations at 353 GHz: in particular, the one-point statistics of $p$ and  $\psi$ in pixel space, and the $EE$, $BB$, and $TE$ power spectra in harmonic space. Our main results can be summarised as follows.

\begin{itemize}

\item The butterfly pattern seen in $q_{\rm d}$ and $u_{\rm d}$ maps (Fig.~\ref{fig:4.2.1}) around the pole is associated with an ordered GMF, which we find has a mean orientation in Galactic coordinates toward $(l_0, b_0) = (62\deg, -18\deg)$ (Sect.~\ref{sec:4.2}). The best-fit value of $l_0$ is roughly consistent with the earlier values derived from starlight polarization \citep{Ellis:1978,Heiles:1996}.

\item From the \planck\ data, we find a $BB/EE$ power ratio of 0.58 and significant positive $TE$ correlation over \rgn\ (Sect.~\ref{sec:4.3}). The key property in our model that allows the model to reproduce these observations (Sect.~\ref{sec:6.3}) is the alignment of the local magnetic field \Bpos\ and the \hi\ structure in the CNM template (Sect.~\ref{sect:3}).

\item  The observed distributions of $p$ and $\psi$ over \rgn\ show a scatter that is comparable to that of the distributions reported for the whole sky in \cite{planck2016-l11B}. Our model successfully reproduced these one-point statistics associated with LOS depolarization (Sects.~\ref{sec:6.1} and \ref{sec:6.2}) by introducing fluctuations in the GMF orientation that are uncorrelated between the three independent \hi\ \phases\ (Sect.~\ref{sec:phases}).

\item The best-fit value of the parameter, $p_0$, that measures the grain alignment efficiency combined with the intrinsic polarization fraction of interstellar dust emission at 353\,GHz, is 15.5\,\%. The best-fit value of the dust emissivity is $\epsilon_{353} = 146\,\mukcmb (10^{20}\,\cm2)^{-1}$ (Sect.~\ref{sec:5.2}). 

\item To match the observed $EE$, $BB$, and $TE$ amplitudes, the model fit yields a low value of the parameter $f_{\rm M}^{\rm u/w}$ specifying the relative amplitude of the turbulent component of the GMF in the  UNM and WNM.  This value, 0.1, is significantly smaller than the value $f_{\rm M}^{\rm c} = 0.5$ characterising the CNM (Sect.~\ref{sec:5.1}). 

\item The spectral index of the turbulent component of the GMF, $\alpha_{\rm M} = -2.4$, fitted over \rgn\ (Sect.~\ref{sec:5.1}), is consistent with the value reported in \tgmodel. These two complementary analyses reveal that the spectral index of turbulent magnetic field closely matches the power-law index of dust polarization power spectra, as expected.

\item Our model also reproduces the inverse relationship between the polarization angle dispersion, $\mathcal{S}$, and the polarization fraction, \pmas, present in the \planck\ data, despite the fact that we do not utilise this phenomenon in fitting the model parameters (Sect.~\ref{sec:7}). Our work reinforces the conclusion in \citet{planck2016-l11B} that the inverse relationship of $\mathcal{S}$ and \pmas\ is a generic feature associated with the GMF structure.

\end{itemize}

This work demonstrates that the phenomenological model introduced in \tgmodel\ for the southern Galactic cap also works well in the northern hemisphere. The next step in our modelling work will be to extend this framework to multiple frequencies by incorporating spectral energy distributions for the dust emission associated with the three \hi\ \phases\, thus introducing the potential for frequency decorrelation of the dust polarization. This is a necessary step towards investigating the utility of this framework for evaluating component separation methods for future CMB missions.

\begin{acknowledgements}

We gratefully acknowledge the use of the Aquila cluster at NISER, Bhubaneswar. DA acknowledges the University Grants Commission India for providing financial support as Senior Research Fellow. 
 This work was supported by the Science and Engineering Research Board, Department of Science and Technology, Govt. of India grant number SERB/ECR/2018/000826 and the Natural Sciences and Engineering Research Council of Canada. 
Some of the results in this paper have been derived using the \healpix\ package. The \planck\ Legacy Archive (PLA) contains all public products originating from the \planck\ mission, and we take the opportunity to thank ESA/\planck\ and the \planck\ collaboration for the same. This work has made use of \hi\ data of the EBHIS survey headed by the Argelander-Institut für Astronomie (PI J. Kerp) in collaboration with the Max-Planck-Institut f\"{u}r Radioastronomie and funded by the Deutsche Forschungsgemeinschaft (grants KE757/7-1 to 7-3). 
The Gaussian decomposition of the EBHIS data was supported by the Estonian Research Council grant IUT26-2, and by the European Regional Development Fund (TK133).

\end{acknowledgements}

\bibliographystyle{aa}
\bibliography{Planck_bib,ebhis,EBHIS_Refs_FB,CPP_L11b,biblio}

\begin{appendix}

\section{Analysis with an alternative set of \hi\ templates} \label{sec:A1}

Here we describe the development of an alternative set of \hi\ templates and then the results from a modelling analysis similar to that presented in the main paper.

\cite{Kalberla:2018} (hereafter KH18) performed a Gaussian decomposition of EBHIS \hi\ emission spectra from an intermediate product of the HI4PI all-sky survey \citep{HI4PI:2016}, applying the algorithm described in \cite{Haud:2000} with some modifications (see their section 2.2). 
Based on line widths, KH18 defined \hi\ components called the cold, luke-warm, and warm neutral medium (CNM, LNM, and WNM, respectively). The separation was done using intervals of line width or the corresponding Doppler temperature ($T_{\rm D}$), which depend on $|v|$ (see, e.g. their figures~3 and 7). For example, for velocities $|v|\, \leq 8$\,\kms\ over the full sky, the mean line widths are 3.6, 9.6, and 23.3~\kms\ FWHM for the CNM, LNM, and WNM components, respectively. 

\begin{figure}[h]
\centering
\includegraphics[width=8.8cm]{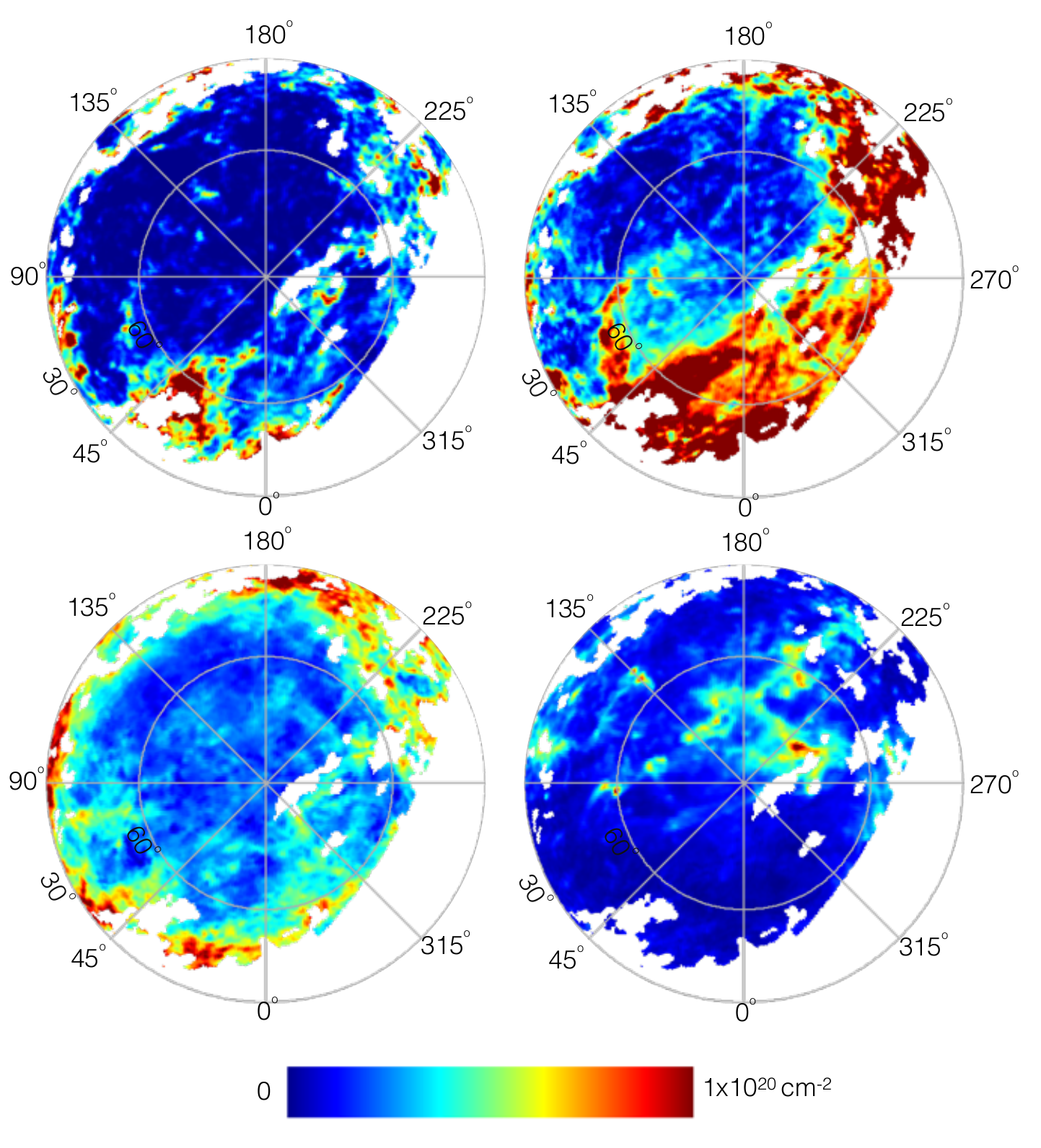} 
\caption{Northern Galactic orthographic projection of \NHI\  of the three KH18 templates, CNM (\textit{upper left}), LNM (\textit{upper right}), and WNM (\textit{lower left}), and the additional fourth template (\textit{lower right}) over our selected region \rgn\ (Sect.~\ref{sec:2.3sec:4}).  The mean column densities of the four templates are $\left\langle N_{\hi}^{\rm c} \right\rangle  = 0.19\times 10^{20} \cm2$, $\left\langle N_{\hi}^{\rm l} \right\rangle  = 0.52 \times 10^{20} \cm2$, $\left\langle N_{\hi}^{\rm w} \right\rangle  = 0.75 \times 10^{20} \cm2$, and $\left\langle N_{\hi}^{\rm 4th} \right\rangle  = 0.37\times 10^{20} \cm2$, respectively. }
\label{figure:A1}
\end{figure}

Making use of this Gaussian decomposition, we consider here Gaussian components whose central peak velocity $v_{\rm c}^i$ lies within the velocity range of $|v_{\rm c}|\, \leq 24$~\kms, which encompasses most of the local \hi\ gas \citep{Kalberla:2018} but ignores most intermediate velocity gas (IVC). This decomposition was done at the native resolution in maps with \Nside = 1024. As in \cite{Kalberla:2018}, maps of the integrated emission of \hi\  were made for the CNM, LNM, and WNM components at an angular resolution of 30\arcm\ (FWHM). For our alternative analysis, these column density maps were further smoothed to 60\arcm\ (FWHM) and downgraded to \Nside = 128. The smoothing reduces uncertainties within each map that arise from separating the gas into components. These maps are shown in Fig.~\ref{figure:A1}. Hereafter, we refer to these as the KH18 templates.

\begin{figure*}[h]
\includegraphics[width=18cm]{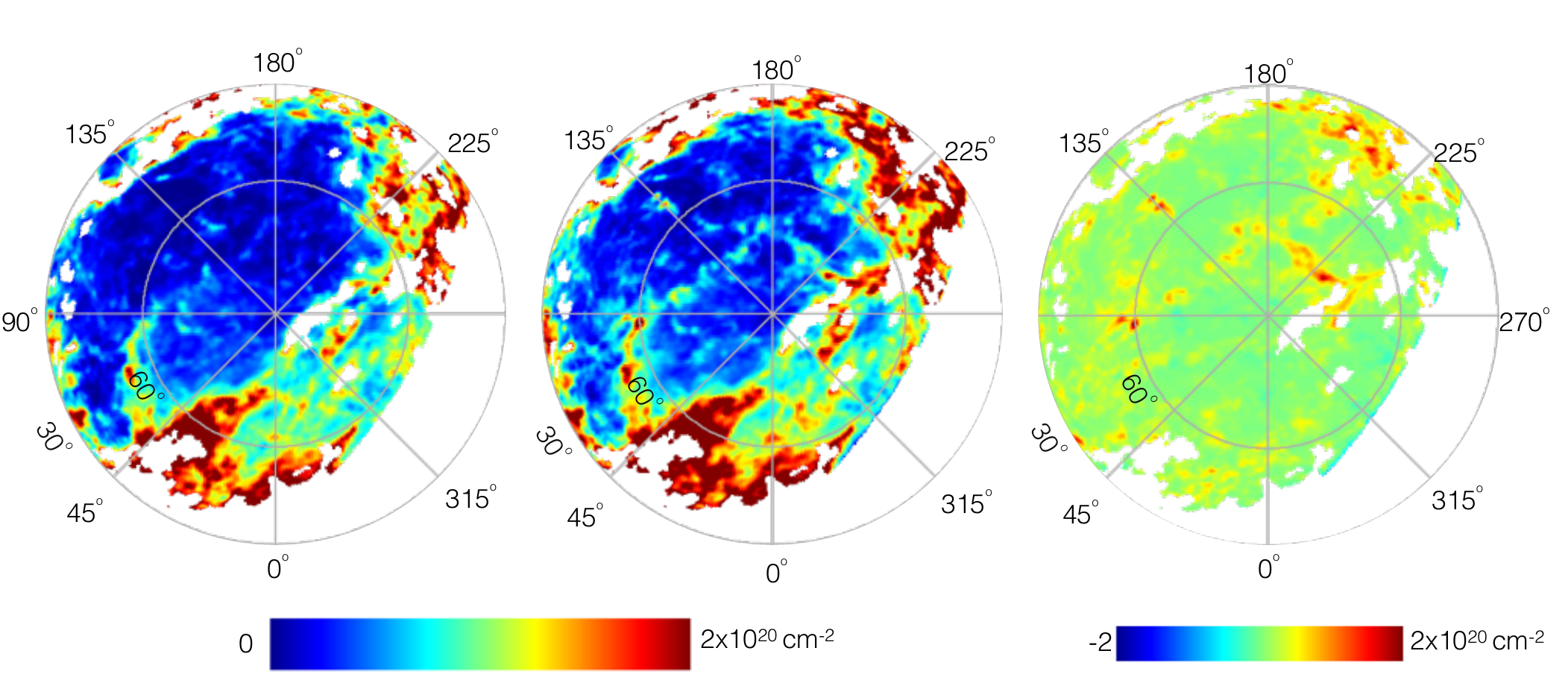} 
\caption{Northern Galactic orthographic projection of our CNM template (\textit{left panel}),  the KH18 CNM plus LNM templates (\textit{middle panel}), and the difference between the two (\textit{right panel}).}
\label{figure:A2}
\end{figure*}

We refer to the sum of the column densities of the three templates as $N_{\hi,24}$. The correlation of dust emission with $N_{\hi,24}$ on the selected region \rgn\ is not satisfactory, having an emissivity $\epsilon_{353,24}$ of $121\,\mukcmb (10^ {20}\,\cm2)^{-1}$, lower than $\epsilon_{353,50}$, and also an unacceptably large offset, $115\,\mukcmb$. This arises because of the restricted velocity range adopted from KH18: $N_{\hi,24} \leq N_{\hi,50}$ and is not an adequate proxy for all of the dust. Therefore, to make the model more rigorous we introduced an additional fourth template to compensate for the \textit{missing} \hi.  The fourth template is simply the column density map $N_{\hi,50}$ - $N_{\hi,24}$, also shown in Fig.~\ref{figure:A1}.

The sum of the KH18 CNM and LNM templates is related to our CNM template used in the main paper, as shown in Fig.~\ref{figure:A2}. The slope of the pixel by pixel correlation between the sum of the KH18 CNM and LNM templates and our CNM template is 1.02 over \rgn, which confirms that CNM and LNM from KH18 can be considered embedded in the gas in our CNM template. Therefore, in building this alternative model we assume that the local \Bpos\ of the model GMF is aligned with \hi\ structures in both of the CNM and LNM templates of KH18 (Sect.~\ref{sect:3}).  

\begin{figure}
\includegraphics[width=9.0cm]{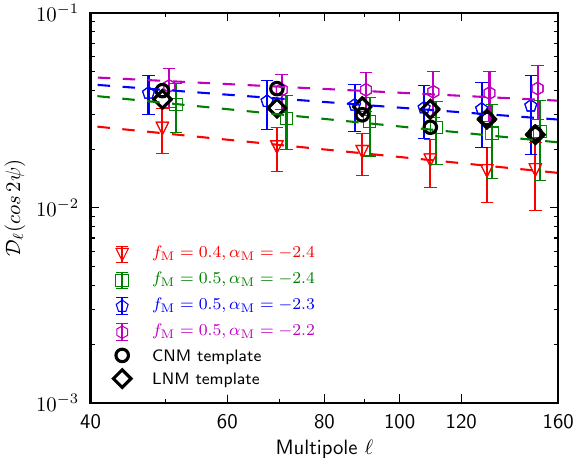}
\caption{Mean of simulated power spectra of $\cos2 \psi$ within the multipole range $40 \leq \ell \leq 160$ for different values of $f_{\rm M}$ and $\alpha_{\rm M}$ and best fit power-law ($\textit{dashed lines}$). The error-bars (1\,$\sigma$) are standard deviations computed from 100 realisations. $\textit{Black circles}$ and $\textit{black diamonds}$ are power spectra of $\cos2 \psi$ computed (Sect.~\ref{sec:phases}) from the CNM and LNM templates of KH18.}
\label{figure:A3}
\end{figure}

Applying the technique discussed in Sect.~\ref{sec:5.1}, we find $f_{\rm M}^{\rm c} = 0.5$ and $\alpha_{\rm M} = -2.4$ (see Fig.~\ref{figure:A3}). Because the error bars on $\cos2 \psi$ estimated from simulated maps are significant, on the basis of Fig.~\ref{figure:A3} we adopt the same pair of parameter values for the LNM. In the WNM, we adopt $f_{\rm M}^{\rm w}$ a lower value, 0.1. The fourth template is assumed to be unpolarized.  For these alternative templates, the other three parameters optimised (Sect.~\ref{sec:5.2}) are
$\epsilon_{353} = 146\,\mukcmb (10^{20}\,\cm2)^{-1}$ and $p_0 = 19$\,\%.

\begin{figure}
\includegraphics[width=9.0cm]{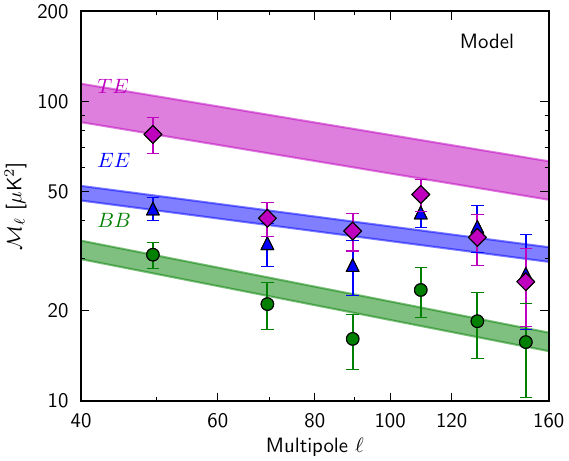}
\caption{Dust $EE,\, BB$, and $TE$ cross-power spectra in units of $\mukcmb^{2}$ computed from the subsets of 100 realisations of the dust model maps over \rgn\ using the  CNM, LNM, and WNM of KH18 and the fourth template. The filled areas represent the \planck\ dust power spectra measurements over \rgn\ computed in Sect.~\ref{sec:4.3}. }
\label{figure:A4}
\end{figure}

Within our multiphase dust modelling framework (Sect.~\ref{sect:3}), these seven parameters and four templates can reproduce (Sect.~\ref{sec:6}) all of the observed statistical properties of dust (Sect.~\ref{sec:4}) quite well, including the power spectra as shown in Fig.~\ref{figure:A4}. 
For $\ell$ bins between 40 and 160, the alternative set of \hi\ templates is able to produce dust $EE$ and $BB$ power spectra with similar strength as the \planck\ data.
The overall  $BB$ power is at the right level and the $BB/EE$ ratio is about 0.57.  

However, the predicted $TE$ is an exception.  There are two concerns.  First, the overall level of the $TE$ power across most scales is low relative to the \Planck\ data (shown originally in Fig.~\ref{fig:5.3.1} left).  Second, the pattern of $\ell$-dependent deviations from a smooth power-law dependence is more pronounced compared to our model in Fig.~\ref{fig:5.3.1}. These deviations could come from excess small-scale power in the KH18 CNM or LNM templates that corrupts the $TE$ (and $EE$) power spectra at small scales.

\end{appendix}

\end{document}